\documentclass[preprint,journal]{vgtc}            %

\newcommand{\fix}[1]{\textcolor{black}{{{#1}}}}
\newcommand{\add}[1]{\textcolor{black}{#1}}
\newcommand{\new}[1]{\textcolor{black}{#1}}
\newcommand{\rev}[1]{\textcolor{black}{#1}}

\onlineid{1060}

\vgtccategory{Research}

\vgtcpapertype{Theoretical \& Empirical}

\title{Multimodal Data Comprehension: Understanding How \\ Visual-Textual Chains of Information Influence Data Interpretation}

\author{%
  Arran Zeyu Wang, Fuling Sun, and Danielle Albers Szafir
}

\authorfooter{
  \item
  	Arran Zeyu Wang and Danielle Albers Szafir are with the University of North Carolina at Chapel Hill (UNC).
  	E-mail: \{zeyuwang, danielle.szafir\} @cs.unc.edu
  \item Fuling Sun is with the University of California San Diego.
    E-mail: fusun@ucsd.edu
}

\abstract{
Visualizations and text often work together to support effective data communication. 
Despite this common paradigm, we know little about how the interplay of these modalities affects people’s data comprehension. 
We present a novel experimental paradigm to investigate \textbf{multimodal data comprehension}---the process of people comprehending information from multimodal visual and textual data---across both crowdsourced and think-aloud environments.
Our methodology employs two sequential chains for presenting multimodal information---a visualization-first chain and a text-first chain---asking people to describe the data presented iteratively.
By comparing how people's data comprehension changes across the chain, we can assess the information contribution of each modality and how they shape subsequent comprehension. We found that the visualization-first chain facilitates \textbf{exploratory comprehension} with hypothesis-driven discovery, whereas the text-first chain yields \textbf{confirmatory comprehension} akin to framing effects where visualizations serve to reinforce and confirm observations drawn from text.
Our findings provide empirical insights into multimodal information integration, with implications for designing more effective data-driven communication.
} %


\teaser{
  \centering
  \includegraphics[width=\linewidth]{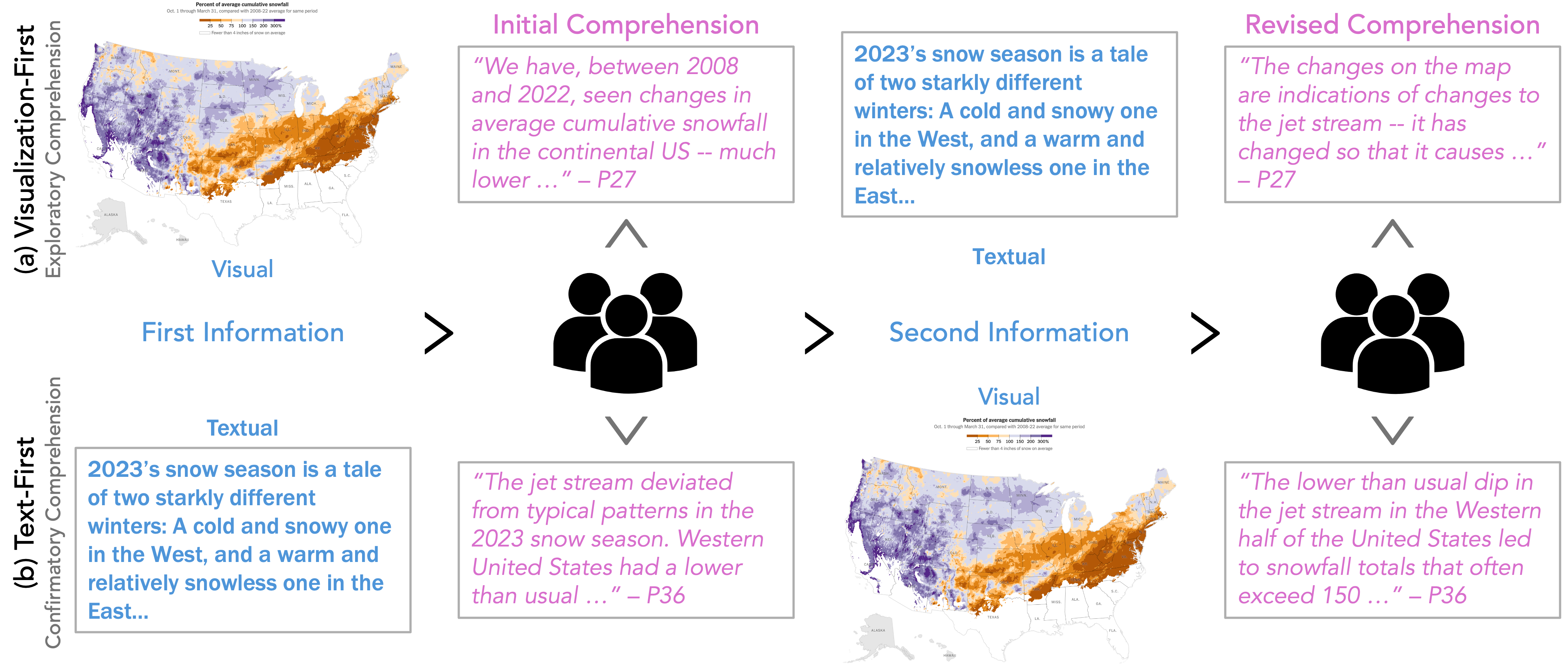}
  \vspace{-0.5em}
  \caption{We investigate how the order of presented information influences people's analytical comprehension and engagement with data.
  The study design compares two sequential information chains: a \textbf{Visualization-First Chain} (top row) and a \textbf{Text-First Chain} (bottom row). Participants viewed these chains and provided open-ended summaries of the data after each step in the chain (pink text). 
  We found that different chains led to distinct patterns in data comprehension.
  (a) The visualization-first chain facilitates an \textbf{exploratory comprehension}, where users move from observing a visual pattern to understanding its cause.
  (b) In contrast, the text-first chain creates a strong framing effect, leading to a \textbf{confirmatory comprehension} where the visualization is used to find evidence for the initial textual narrative.
  This difference illustrates how chains of information can be used to either encourage discovery or guide an audience to a specific conclusion.
}
  \vspace{-0.5em}
  \label{fig:teaser}
}

\graphicspath{{figs/}{figures/}{pictures/}{images/}{./}} %

\usepackage{tabu}                      %
\usepackage{booktabs}                  %
\usepackage{lipsum}                    %
\usepackage{mwe}                       %

\usepackage{mathptmx}                  %
\usepackage{amsmath,amsfonts}
\usepackage{algorithmic}
\usepackage{algorithm}
\usepackage{array}
\usepackage{textcomp}
\usepackage{stfloats}
\usepackage{url}
\usepackage{verbatim}
\usepackage{graphicx}
\usepackage{cite}  
\usepackage{amsmath}

\begin{document}

\maketitle

\section{Introduction}
\label{sec-intro}

From academic papers to public-facing dashboards, visualizations are relied upon to efficiently convey information and translate complex data into actionable insights~\cite{franconeri2021science, gotz2008characterizing, segel2010narrative}.
However, \add{in practice, visualizations are rarely consumed in isolation.} 
Instead, they are frequently used \add{as part of a multimodal presentation, combined with text (e.g., a news article~\cite{rogha2024impact}), imagery (e.g., a data comic~\cite{bach2017emerging}), or audio (e.g., a documentary~\cite{bradbury2020documentary} or oral presentation~\cite{stokes2024delays}).}
Such context, often provided through captions, titles, or surrounding narrative text, plays an essential role in guiding interpretation and shaping the reader's comprehension~\cite{segel2010narrative, stokes2023role, quadri2024do}.

However, \add{the critical interplay between text and visuals 
is 
poorly understood---visualization research often examines charts in isolation~\cite{hearst2023show}.}
Graphical perception and related studies characterize effectiveness largely based on visual design, failing to consider textual context~\cite{szafir2023visualization, cleveland1984graphical, wang2025characterizing}.
Prior work has explored the impact of text on how people read visualizations, but primarily focused on simple, descriptive information, like titles, annotations, or standalone statistics and statements~\cite{stokes2022striking, stokes2023role, borkin2015beyond, kong2018frames, rahman2025survey}.
However, these techniques may be essentially re-explaining the chart, often providing information people could have inferred directly from the visualizations~\cite{lundgard2021accessible, quadri2024do, tang2023vistext, stokes2025analysis}.
\add{In real world contexts like 
news articles~\cite{theNYTgraph}, text serves a richer narrative function, providing context, externalized knowledge, and expert opinion that is not immediately visible in the chart itself.}
We lack insight into how rich multimodal information in real-world communication \add{shapes data comprehension} (see  \autoref{fig:teaser} for \add{examples}).
In this work, we 
investigate how people comprehend data across visual and textual information 
under such conditions.

\add{A key factor that may impact the interplay between text and visualizations in data communication is the \emph{order} in which information is consumed.}
\rev{In many common daily usage scenarios}, people consume information sequentially, scanning their attention from text to visualization or visualization to text.
Imagine reading a news article about climate change: if you see a shocking graph of rising temperatures first, you may likely feel alarmed and engage with the data to hypothesize about the cause of the drastic change before reading the text for explanations. 
\add{Conversely}, reading a fearful headline first, "Climate disaster imminent...", may lead you to see the same graph merely as proof of that disaster, not as a source for exploration or investigation. In other words, 
\add{you may be primed to look at data as evidence for a specific conclusion rather than as a means for exploration and critical engagement.}
This difference would mean the order in which you receive the information can change \add{
how you comprehend the same data. Seeing the visualization first may cause you to engage with and actively process the data, attending to and interpreting different patterns using sensemaking processes, whereas text first may prime you to directly search for confirmatory evidence.}

\new{The order of information delivery can commonly change in the real world. For example, if you read a news article or paper from a website or mobile phone, it is common to see teaser images first, but if you receive a breaking news alert first through your email or listen to a podcast when driving, text becomes lead modality and the information chain is reversed.
Other factors, such as the exact article or paper layout
and 
seeing text summaries ahead of graphics in an LLM response, may likewise shift the order of information.}

\add{We investigate the role of order in multimodal data communication}
through a novel bidirectional experimental paradigm investigating how the \add{order} of \add{visual and textual information} shapes people's evolving understanding of data. 
We examined two fundamental 
chains of visual and textual information 
common in real-world data communication.
In the \textit{\add{visualization-first} chain} (\autoref{fig:teaser} (a)), participants first view a data visualization and write an initial comprehension reflecting their understanding of the data, then receive text information and report their revised comprehension.
In the \textit{\add{text-first} chain} (\autoref{fig:teaser} (b)), this order is reversed: participants see the textual information first, followed by the visualization.
By assessing initial and revised comprehensions within and between these chains, we can empirically assess information transmission across modalities and changes in comprehension provided by each modality.

We applied this approach using visual graphics and stories from \textit{The New York Times}~\cite{theNYTgraph} which contains rich visual and textual information written and designed by professional journalists.
We selected stimuli consisting of multiple paragraphs of article text with accompanying visualizations
and assessed participant's comprehensions \add{with two chains in both crowdsourced ($N=60$) and in-person think-aloud ($N=8$) environments}.
Our results provide empirical insights into the unique and complementary roles of visual and textual information in sensemaking (\autoref{fig:teaser}).
Notably, in the \add{visualization-first} chain, people's comprehension initially describes a diverse set of 
observations, visual patterns, and hypotheses, exhibiting \textit{exploratory} behavior and using the text to refine and reason causally about their observations.
In the \add{text-first} chain, people start with a clear, unified, but also restricted understanding of the text and engage in \textit{confirmatory} behaviors when encountering the visualization.
\add{Seeing subsequent charts has little impact on their expressed data comprehension 
as the visualization becomes supporting evidence that the text is ``correct.'' }
Our findings provide a framework for understanding how people comprehend multimodal data with different chains of information and suggest implications 
for more effective and intuitive multimodal communication 
approaches.
Our work moves beyond seeing text as a supplemental annotation or label, treating it as an equal partner in the construction of knowledge in visualized data \cite{hearst2023show} and revealing how this partnership can shape data sensemaking. 
The main contributions of this paper include:
\begin{itemize}
    \item 
    A 
    methodology for empirically studying multimodal information integration in data comprehension.
    \item Empirical evidence that the chains of information in multimodal data affect comprehension, creating two distinct behaviors: 
    either clarifying and extending initial explorations (\textbf{exploratory} comprehension) or making a prior understanding more concrete (\textbf{confirmatory} comprehension).
    \item Our findings offer actionable guidance for designers on how to 
    organize visual and textual information to optimize data communication to either encourage open discovery or ensure accurate and efficient comprehension. 
\end{itemize}

\section{Background}
\label{sec-related}

Our work 
draws conceptual inspiration from information theory and cognitive science and builds upon 
research in data storytelling, visual comprehension, and text-based communication.

\subsection{Effective Visual Communication: Storytelling with Data}

Beyond purely analytical tasks, 
visualizations are often used for data storytelling~\cite{franconeri2021science}.
Effective
storytelling 
designs thoughtfully direct a reader's attention through a narrative~\cite{knaflic2015storytelling}.
The rhetoric~\cite{hullman2011visualization} and the presentation order of 
visualizations~\cite{hullman2013deeper} influence the reader's interpretation of a data story.
Visual narrative flows, such as visuals and navigation feedback, help control the order in which people encounter different visualizations to influence people's understanding and engagement in a story~\cite{mckenna2017visual}.
These works highlight that data communication and storytelling involve complex design processes to optimize across a space of design guidelines and heuristics to get the desired communication performance.

Modern news media and journalism 
often draw on visual storytelling theories and best practices to create 
articles and reports~\cite{wahl2019news, kulkarni2023innovating}.
Visualization research has suggested different practices to help people architect data stories, such as leveraging narrative guides like \emph{Freytag's Pyramid} \cite{yang2021design}, supporting automated linking of text and charts \cite{sultanum2021leveraging}, or conceptualizing frameworks for crafting non-traditional storyflows \cite{chen2018supporting,bach2017emerging,hao2024design}. Much of this past work focuses on mechanisms for designing and implementing stories; however, the structure of a data story can impact the reader's experience as well. 
Segel \& Heer~\cite{segel2010narrative} identified common narrative structures used in journalistic and explanatory visualizations, such as the \emph{martini glass} structure~\cite{boy2015storytelling} that first presents an author-driven narrative and then allows for reader-driven exploration.
Besides, the structure of a visual story~\cite{zhi2019linking} as well as its temporal or non-temporal order~\cite{lan2021understanding} may also influence comprehension and engagement.
These studies investigated how different genres of story structure influence interpretation; however, they lack insight into more granular components of story design, including how knowledge evolves over the course of a story or across modalities and how that evolution may shape readers' information foraging behavior. 
We probe the progress happened across communication modalities in 
data stories by assessing how its two primary components---visuals and words---work together to 
shape people's understanding of data.

\subsection{High Level Data Visualization Comprehension}

Research has long studied how visualizations support insight into data (see Quadri \& Rosen~\cite{quadri2021survey} and Battle \& Ottley~\cite{battle2023we} for surveys).
Seminal work in visualization perception and cognitive science established the foundations for understanding how people decode visual encodings and identify patterns~\cite{cleveland1984graphical}
and introduced frameworks for how people reason with data \cite{russell1993cost}.
This research has informed 
design principles for effective visualization, from 
selecting appropriate chart types~\cite{ware2012information} to 
using color~\cite{szafir2018modeling}, shape~\cite{tseng2024shape}, and redundancy~\cite{tseng2026redundant}.

Most recently, empirical and theoretical research in visualization comprehension has evolved from studying low-level perceptual tasks~\cite{amar2005low} like value estimation to understanding cognitive psychology processes~\cite{wang2025characterizing}, including complex information comprehension~\cite{quadri2024do, bearfield2024same, wang2026contextualization}, cognitive biases~\cite{li2025confirmation, wang2024causal}, and decision making~\cite{padilla2018decision, kandel2025graphical}.
For example, existing work found people often extract gist level information rather than specific point values from bar and line graphs~\cite{shah2011bar}. People's high-level comprehension of a visualization 
is frequently misaligned with designer's objectives~\cite{quadri2024do} and is influenced by their prior beliefs and expectations~\cite{kim2020bayesian}. Even the same data representation can yield different mental models and understandings~\cite{williams2023data, bearfield2024same, wang2026contextualization}.

These studies typically examine visualizations in isolation.
Yet, in many common real-world contexts, visualizations 
are accompanied by text, annotations, and other complementary modalities that enrich the data to support comprehension and sensemaking. 
Further, people attend to information and build knowledge from that information sequentially, even within a singular visualization. Elements of this context, such as external emphasis \cite{xiong2019curse}, may shape the order in which people attend to different information. While frameworks like sensemaking \cite{pirolli2005sensemaking} can model the high-level process of reasoning with visualizations, we have little insight into how knowledge is constructed from visualizations in richer contexts where data is often presented across different modalities and how comprehension evolves across modalities and over time.

\subsection{The Role of Text in Visualization Communication}
Visualizations are rarely consumed in isolation, they are also accompanied by other content, such as video, audio, animation, and text.
Studies examining the effects of 
such multimodal content have investigated the text scaffolding explaining information shown within the chart itself,
exploring how titles, labels, and annotations affect chart reading speed, recall, and accuracy~\cite{stokes2025analysis}.
For instance, 
a descriptive title can help prime people to identify key insights more quickly~\cite{wanzer2021role}, but a misaligned one may increase bias~\cite{kong2019trust} and cause incorrect recall \cite{kong2018frames}.
Well-crafted annotations improve memorability~\cite{borkin2013makes}, recall~\cite{borkin2015beyond}, and uncertainty interpretation and reasoning~\cite{rahman2024qualitative}, and may also influence
bias and data-driven predictions~\cite{stokes2023role}.
Integrating information-rich captions with charts can significantly enhance people's ability to accurately interpret information
and clarify ambiguities~\cite{stokes2022striking, stokes2024give}.

These studies indicate that text is important for narrative construction and sensemaking.
However, they predominantly treat text as a secondary component, an aid to help interpret the primary visual presentations. In practice, text often serves as the predominant means for communicating data, with visualizations 
augmenting the text-based content of a narrative rather than serving as the primary mechanism for information delivery \cite{segel2010narrative, stolper2018data, stokes2024give}. Hearst argues for treating language as co-equal to visualization in research \cite{hearst2023show}, noting a variety of ways in which text and visualization may interact~\cite{stokes2024give}. 
We build on this perspective, treating text not as an augmentation in service of a visualization but as an independent source of information that 
works directly with visualizations and other data communication modalities. These modalities are then processed by cognitive systems to help people reason over and comprehend data. 
Our approach aims to 
explore the behaviors resulting from this process. %

\subsection{Sequential Information Processing of Cognition}

People typically do not process information at random: 
we attend to information sequentially, and 
the order of exposure matters~\cite{friedman2001memory}.
Our treatment of data and text as information chains is inspired by classical communication paradigms that illustrate how information can be gained or lost in a sequential process~\cite{fisher1984narration, lashley1951problem}.
First formally discussed by Lashley as \emph{priming}~\cite{lashley1951problem}, further foundational work from Kahneman \& Tversky in cognitive psychology has established strong theoretical basis for sequential phenomena---the \emph{framing effect}~\cite{tversky1989rational}, where initial stimuli influence the response to subsequent stimuli, and the \textit{anchoring and adjustment heuristic}, where an initial piece of information serves as an anchor that biases future judgments~\cite{tversky1974judgment}.
Such sequential paradigms have been used to study information propagation and the fidelity of message transfer in human groups~\cite{lerman2016information, vlasceanu2018cognition, stone2019individual}.

These effects 
exist in visualizations as well. 
Experiments demonstrate that seeing a scatterplot with different visual patterns will change people's perception of class separability in the next scatterplot~\cite{valdez2017priming}.
Further, priming and anchoring effects 
affect how people characterize clustering in multiview scatterplots 
\cite{zhou2024examination}.
Established beliefs about data can change people's estimation of correlation~\cite{xiong2022seeing} and causality~\cite{wang2026contextualization} for the same visualization with different labels.
Several factors, such as a person's locus of control, may impact people's susceptibility to anchoring and priming in visualization~\cite{alves2024studying}.
Framing effects and sequentially adding and omitting information from visualizations were found to impact to people's interpretation in the context of narrative visualizations~\cite{hullman2011visualization, hullman2013deeper}.
In more complex visual analytics contexts, 
the order and magnitude of information shown in the analytics pipeline can influence decision making
\cite{cho2017anchoring}
and analytical focus during exploration~\cite{zhou2021modeling}.

However, established studies on framing and anchoring effects in visualization focus largely on exploratory analytical pathways and the effect one visualization has on another without considering complementary modalities.
Our study adapts sequential information presentation methodologies in a 
two-stage format to 
understand how comprehension evolves in sequence depending on the order in which different modalities are delivered.

\section{Methodology}
\label{sec-method}

\new{To understand the influence of information order on data comprehension in multimodal comprehension, we introduce a paradigm centered on multimodal communication chains.
This approach allows us to capture and compare a reader's evolving understanding of data across modalities of information.}
The study was approved by UNC Institutional Review Boards.

\subsection{Materials and Stimuli}

We tested comprehension using 12 stimuli sourced from 
stories published by \textit{The New York Times}~\cite{theNYTgraph} that used multimodal data presentations.
We chose this source for its high ecological validity, as the materials were created by expert journalists and visual designers for broad audiences 
following established best practices for news media~\cite{ng2014new}.

Each stimulus consisted of a data visualization and its accompanying multi-paragraph text
article. 
We manually reviewed 181 news articles including visual graphics published from 2020 to 2024 (see \autoref{tab:chartall}
for a summary), and selected 12 graphics 
that research team agreed covered a diverse range of daily visualization usage scenarios, contained sufficient visual and textual information to draw meaningful insight from each modality independently, and were not too complex for lay readers. 
Stimuli consisted of one visualization and two to three paragraphs from each article
communicating messages related to the core ideas in the associated visualization,
with word counts ranging from 80 to 147 words \new{from 12 news articles}.

We then further reduced the corpus to select visualizations reflecting the information density, diversity, and authenticity of stimuli found in real-world data communication.
We assigned labels of chart type and topic \new{(e.g., energy, framing, health, etc.)} for each visual graphic, and applied five selection criteria to choose the final 12 stimuli to ensure familiarity, clarity, representation, neutral attribution, and topic diversity. 
\new{See \autoref{tab:chart} for the distribution of selected visualizations and \autoref{fig:teaser} and \autoref{sec-eval} for example stimuli.}
Please see \autoref{appd-stimuli}
for additional details on stimulus curation, including the selection criteria, process, visualizations, and topics, as well as all stimuli (\autoref{appd-all-stimuli}).

\begin{table}[htbp]
\renewcommand{\arraystretch}{1.2}
\vspace{-0.5em}
\centering
\caption{Stimulus Visualization Type Distribution}
\vspace{-0.5em}
\label{tab:chart}
\scalebox{0.92}{
\begin{tabular}{|c|c|c|c|c|c|}
\hline
Types & Map & Line Chart & Bar Chart & Area Chart & Scatterplot \\
\hline
Counts     & 3    & 3  & 3   & 2   & 1       \\
\hline
\end{tabular}
}
\vspace{-1.5em}
\end{table}

\subsection{Study Design}
We employed a within-subjects design where the primary independent variable was the \textit{Information Presentation Order}.
This variable had two levels, one for each communication chain: 

\noindent \add{\textbf{The Visualization-First Chain}:} Participants were first shown the visualization in isolation and were then presented with the accompanying text (\autoref{fig:teaser} (a)).

\noindent \add{\textbf{The Text-First Chain}:} Participants first read the paragraphs in isolation and were then shown the corresponding visualization (\autoref{fig:teaser} (b)).

\add{We conducted the study across both crowdsourced and in-person participants. The crowdsourced participants allowed access to a larger and more diverse sample population \cite{douglas2023data}, while the in-person participants allowed us to directly observe participants' analytical behavior through thinkaloud methods and provided greater control over data quality.}
We captured participants' data comprehension after each step in the chain, using open-ended free-form text descriptions \add{for crowdsourced participants} \add{or thinkaloud verbal responses for in-person participants}.

\add{For the crowdsourced experiment}, each participant was randomly assigned 4 of the 12 stimuli:
two using the visualization-first chain and two
using the text-first chain, with the mapping of the communication chain and stimuli counterbalanced between participants. \add{We counterbalanced chains and stimuli across participants and randomized the order in which people saw each of the four trials to mitigate potential transfer effects.}
To prevent learning effects from specific visual encodings, we also ensured that no participant saw the same chart type more than once across their four trials, and no participant saw the same stimuli in both a visualization-first and text-first chain.
\add{Completing only four stimuli helped avoid fatigue effects that can degrade data quality in crowdsourced studies \cite{hossfeld2014best}. We opted to include all twelve stimuli in our samples to increase stimulus diversity and help account for individual variation between participants in our final dataset. }

\add{For the think-aloud experiment, each participant 
completed all 12 stimuli. Given the nature of in-person studies, twelve stimuli proved manageable without significant fatigue or other transfer effects in piloting.
Similar to the crowdsourced study, each participant evaluated six stimuli using the visualization-first chain and six stimuli using the text-first chain, with the order of the twelve stimuli and mapping between stimuli and chains both randomized.
In the study, participants were encouraged to think-aloud to help capture behaviors not accessible in crowdsourced studies (e.g., reasoning processes or affective responses).}

\subsection{Procedure}
The overall procedure was the same for both populations. Participants began the study by first providing informed consent and reading brief instructions summarizing the experimental task. Participants then completed a brief demographic questionnaire and entered the formal study.

Participants saw the sampled trials (four for crowdsourced and twelve for in-person) in a random order.
For each trial, participants first 
saw the first modality in the chain and provide their initial comprehension of the data 
based on this modality.
They then advanced to the second modality and provided a revised comprehension. This process proceeded for each chain as follows:

\vspace{3pt}
\noindent Visualization-First Chains:

    \noindent \textbf{Initial Comprehension}:
    The participant was shown only the data visualization and prompted:
    ``\textit{Based on this chart, please describe your comprehension of this data.}''
    
    \noindent \textbf{Revised Comprehension}:
    After submitting their initial description, participants were shown the accompanying text without the image
    and asked for an updated response: 
    ``\textit{Combining the above paragraphs and the previous chart, please describe your comprehension of this data.}''

\vspace{3pt}
\noindent Text-First Chains:

    \noindent \textbf{Initial Comprehension}:
    The participant was shown only the text of the article and prompted: ``\textit{Based on the above paragraphs, please describe your comprehension of this data.}''
    
    \noindent \textbf{Revised Comprehension}:
    After submitting their response, they were shown the corresponding visualization without the text and
    asked to provide a revised description: ``\textit{Combining the above chart and the previous paragraphs, please describe your comprehension of this data.}''

\vspace{3pt}
Crowdsourced participants provided their comprehensions using a text box, while in-person participants were encouraged to verbalize their reasoning processes as they constructed the comprehensions and then provide a verbal summary comprehension. 
This process was repeated for all assigned stimuli.

After completing the formal trials, all participants were asked to describe their comprehension strategies when the data were presented in the two different chains and their overall experience during the study. \add{Crowdsourced participants provided these responses through a textbox while in-person participants responded in a semi-structured interview.}

\subsection{Participants}

We recruited \add{crowdsourced} participants from Prolific~\cite{palan2018prolific}.
All participants had normal vision or corrected-to-normal vision, and provided informed consent before beginning the study.
\add{Three engagement checks were randomly placed among the questions (e.g., select the answer of 17 plus 3). If a person selected the wrong answer, the study was terminated and 
a new participant was recruited.}
60 \add{qualified} participants (36 male, 23 female, one non-binary) were involved in the crowdsourced study.
Among them, each participant took an average of \add{21 minutes 14 seconds} to complete the study.
In-person participants 
were recruited through mailing lists, professional networks, community outreach, and snowball sampling.
Eight users were involved in the study (four male, four female).
All participants had at least a university degree and self-reported some familiarity with visualization.
The thinkaloud study took about 1 hour.
\rev{Our study has 14 responses per stimulus/condition on average, fitting the common sample size in HCI research~\cite{Caine2016local}.}

\subsection{Data Collection \& Analysis}

The primary data collected from the experiment were the text descriptions provided by the participants from both studies.
In total, we collected 480 comprehension descriptions (60 participants $\times$ 4 stimuli $\times$ 2 comprehensions per stimulus) from crowdsourced study and 192 descriptions (8 participants $\times$ 12 stimuli $\times$ 2 comprehensions per stimulus) from thinkaloud study.

We analyzed the data using thematic analysis.
Two coders collectively identified, discussed, and categorized codes using an inductive approach and summarized the main themes from the resulting codes.
Coders met regularly throughout the process to assess agreement
and refine the codebook accordingly.
Discrepancies were discussed until consensus was reached.
Each coder individually coded half of the corpus plus 40 overlapping responses selected at random to verify interrater reliability.
We computed Cohen's kappa $k=0.85$, indicating excellent reliability~\cite{mchugh2012interrater}.
\add{We extracted themes from the coded comprehensions, with an emphasis on 
observed comprehension patterns emerging between initial and revised comprehensions (i.e., \textbf{exploratory comprehension} and \textbf{confirmatory comprehension}) that reflect the change in comprehension from information ordering.}

These codes and analysis procedures enabled us to comprehensively capture critical patterns in people's comprehension and allow a global comparison between their comprehension across modality, chain direction, and stimulus.
Our discussion of results focuses on the primary findings from our analysis.

\section{Findings}
\label{sec-eval}
We found that how people comprehend multimodal data is driven by the chain of information presentation.
The order in which a reader consumes visual and textual information not only changes their final conclusions from the data, but 
may alter their sensemaking process. 
We observed 
two distinct patterns of comprehension correlated with presentation order: 
\textbf{exploratory comprehension} in the \add{visualization-first} chain (\autoref{sec-res-for}) and \textbf{confirmatory comprehension} in the \add{text-first} chain (\autoref{sec-res-back}). %
We additionally synthesized 
a cognitive framework of how people consume data across 
different multimodal information chains based on patterns in the coded data (\autoref{sec-res-behavior}).
Participant quotes from the \add{crowdsourced} experiment are denoted by a randomly assigned an ID from $P1$ to $P60$.
\rev{We use P to indicate participant counts (e.g., 60/60 P) and R forresponse counts (e.g., 120/120 R).}

\subsection{\add{Visualization-First} Chain: From Visual Exploration to Narrative Reasoning}
\label{sec-res-for}

When participants viewed the visualization first, their comprehension process was characterized by an exploratory and evidence-driven approach.
Their initial descriptions of the data began with concrete visual patterns
while the text enabled a richer contextualized understanding embodied in the revised comprehensions as explained 
hypotheses and 
causal reasoning.

\subsubsection{Initial Comprehension (Visualization Only): Diverse Pattern-Seeking and Hypothesis Generation}

\begin{figure}[tb]
    \centering
    \includegraphics[width=\linewidth]{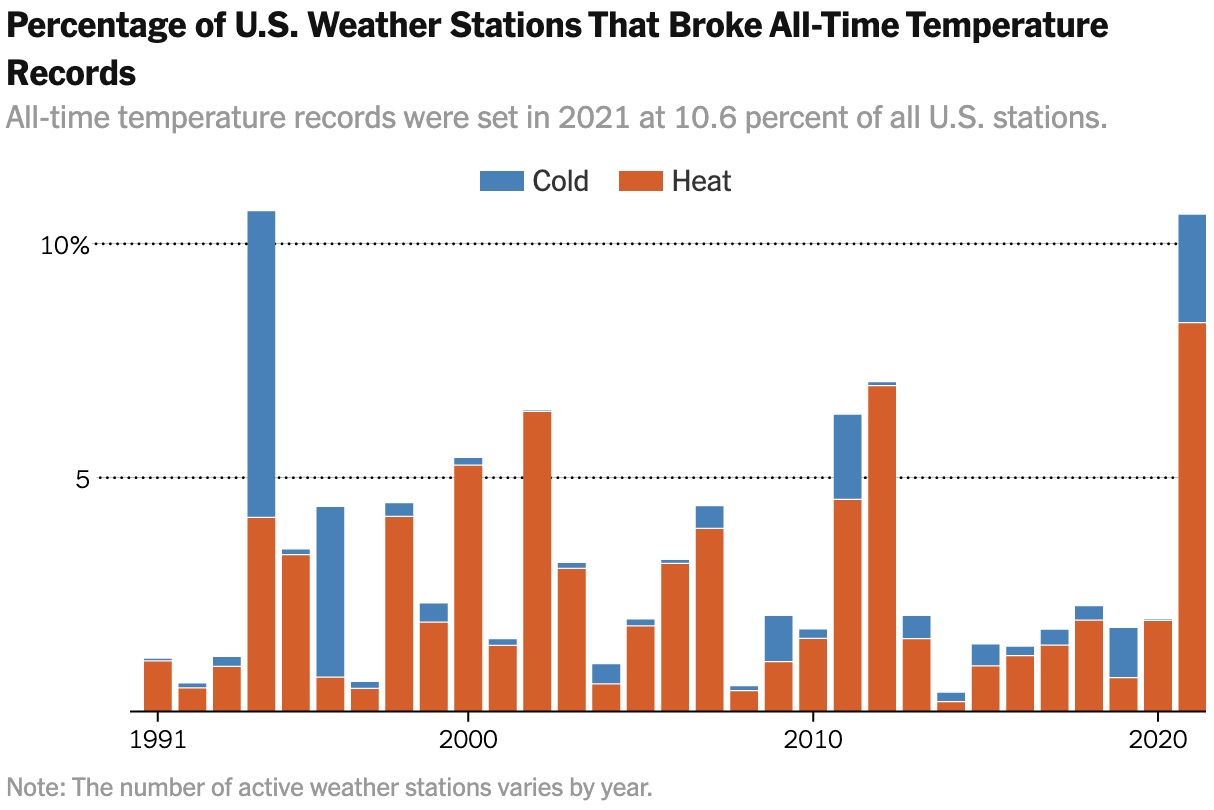}
\vspace{-1em}
    \caption{The visual graphic of US Temperature Record~\cite{NYT1}.}
    \label{fig:ustemp}
\vspace{-1em}
\end{figure}

Viewing data with only the visual representations, participants' responses provide evidence of active exploration.
Their initial descriptions mainly focused on 
salient statistical visual features (101/120 R), such as areas/ranges, trends, outliers/extremes, clusters, and other patterns.
For example, after seeing the stimulus of US Temperature Record (see \autoref{fig:ustemp}), \add{most participants reported visual statistical features (9/10 P)}.
$P2$ noted, ``\textit{The chart shows that U.S. weather stations have consistently broken more heat records over the past three decades...}''
describing a continuous trend and situated that trend within the chart's context.
\new{$P10$ expressed a similar comprehension for the same map, stating that ``\emph{Weather is changing, most new records are recording as hotter than previous records continuously.}''}
This emphasis on different statistics aligns with past studies on visualization comprehension that found people comprehend a range of different statistical quantities in visualized data~\cite{quadri2024do}.

Crucially, this visual exploration was often coupled with spontaneous hypothesis generation (67/120 R).
Without textual context, participants actively speculated on the underlying causes of the patterns they observed, often posing their thoughts as questions.
In the US Temperature example above,
$P3$'s comprehension preliminarily relied on an assumption that the trends observed were periodic: 
``\emph{temperature records go back and forth ... in almost a cycle like pattern}. 
In 
a chart describing farming and underground water in Kansas (see \autoref{fig:farmwater}), half of the participants  
observed a potentially causal correlation: ``\emph{As water levels fall so does corn yield...}'' [$P25$], describing how falling water impacts declining corn yields.
Descriptions were sometimes also filled with uncertain but curious language. For example, two also primarily described the correlation pattern between water and corn, but reflected a more uncertain conclusion of ``\emph{indirect correlation}'' that is ``\emph{not completely conclusive}'' [$P24$].
This pattern of digging into uncertain yet causal reasoning reflects an experience of curiosity
and a desire to 
understand the meaning behind the information presented by the chart.

The same visual graphic can lead to a range of diverse comprehensions that focus on different patterns, further reflecting visualization's power to elicit different understandings~\cite{bearfield2024same, quadri2024do}. \new{ On average, one visual graphic can lead to three to four types of different focuses in people's interpretations}.
In the US Temperature Record example (\autoref{fig:ustemp}), \add{participants reported five different statistical patterns, and no single pattern was universally described across all participants.}
\add{For example, rather than focusing on overall patterns like $P3$'s periodic observations described above (also noted by 4 of 10 participants)}, some participants \add{(3/10 P)} observed an extreme pattern at a peak point of temperature in the chart, referencing that specific statistic in their comprehension as in ``\emph{the number of temperature records that were broken in 2021...}'' [$P4$].
Others \add{(2/10 P)} read both the highest and the lowest to interpret the data aggregated by year, as
    ``\emph{in 2021 ... all-time temperature records ... The highest number of all-time cold temperatures was in 1994.}'' [$P7$].

The Farming and Underground Water example (\autoref{fig:farmwater})
shows similar variance in the patterns people use to describe data. Rather than emphasize correlations \add{(5/10 P)} like $P24$ and $P25$ above, others \add{(3/10 P)} focused on an unusual change of the data pattern from two different time periods, such as,
\begin{quote}
    ``\emph{Even though water levels dropped from 1960 to 1980, corn yields doubled.  But, beginning in the year 2000, as water levels dropped, corn yields also dropped significantly}'' [$P27$].
\end{quote}
On the other hand, $P29$ interpreted the visualization with a focus on temporal patterns on how values change within specific ranges,
\begin{quote}
    ``\emph{In 1960, the water level was around -100 feet below land surface and as the years progressed, it fell. In 1980 for instance, it had fallen to -130 feet below land surface and then got to -180 feet below land surface in 2020.}'' [$P29$].
\end{quote}
These observations show that people draw a diverse array of conclusions even from the same chart. Of note, this diversity occurs despite the fact that the tested visualizations were typically designed to communicate one specific pattern in the data to support the source article's narrative.

\subsubsection{Revised Comprehension (Visualization + Text): Hypotheses Validation, Causal Reasoning, and \fix{``\emph{Aha}'' Moment}}

Viewing the text after the visualization led to comprehensions that synthesized observations across patterns and 
imbued additional meaning into the visualizations.
Participants (91/120 R) commonly used the textual narrative to explicitly validate their discoveries, confirm or correct their initial hypotheses, and generate causal inferences, adding qualitative and contextual information derived from the text.
For example, after seeing the textual information of the US Temperature Record (\autoref{fig:ustemp}), \add{many participants (6/10 P) included language validating
their previous findings or hypotheses}. 
$P2$, who had previously reported a monotonic trend of temperature, further confirmed this finding in their revised description and added more nuance to their initial observations, 
\begin{quote}
    ``\textit{...the text explains the depth and reliability of the data...From both sources, it becomes evident that all-time records are a yearly occurrence, but the scale and intensity vary...}'' [$P2$].
\end{quote}
Similarly, $P10$ confirmed the ``\emph{likely}'' pattern from their initial comprehension (cyclic behavior) and extended the their conclusions based on causal observations to make predictions based on the combined information, explaining that ``\emph{our climate is changing and going towards much warmer weather.}''

\begin{figure}[tb]
    \centering
    \includegraphics[width=\linewidth]{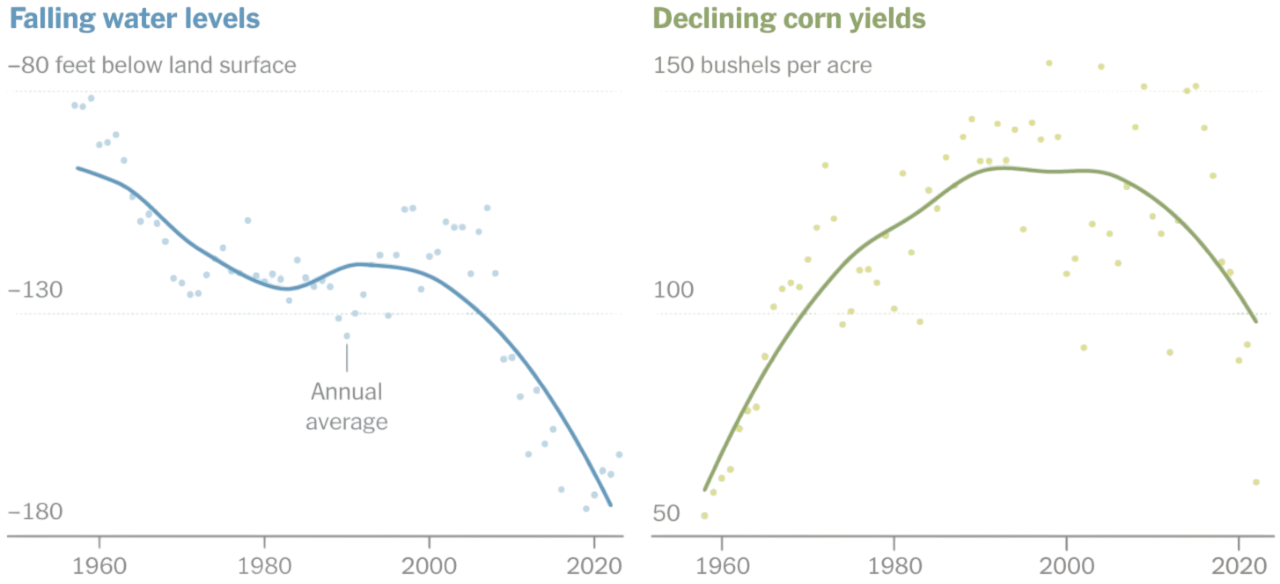}
\vspace{-1em}
    \caption{The visual graphic of Farming and Underground Water in Kansas~\cite{NYT5}.}
    \label{fig:farmwater}
\vspace{-1em}
\end{figure}

This revision process often involved a targeted re-examination of the visualization, where people would use causal inferences drawn from the text to further reason about their chart-based comprehension. 
In the Farming and Underground Water example (\autoref{fig:farmwater}), \add{7 of 10 participants exhibited this behavior}. $P25$, who initially described the data as showing a correlation between water and corn yields, revised their initial purely statistical description to use firmer and more precise causal language:
``\emph{The water being so scarce is causing fields to dry out and corn cannot produce [sic] in these conditions.}''
The final revised comprehension represented a grounded, causal narrative, using the 
textual story to form a high-level takeaway to explain the statistical evidence drawn from the visualization.

While the above examples contextualized individual patterns, text was also used to create causal hypotheses about the relationship between multiple patterns. 
For example, $P26$ connected two disjoint patterns from their original interpretation of the Farming example
using context from the text:
    ``\emph{This text provides context to explain the relationship between the two charts. From 1960-2000, corn yields increased because farmers were draining the aquifer. After 2000, they only relied on rain, which decreased yields.}''

\add{As seen above, seeing the text after the visualization often introduced causal reasoning 
into participants' descriptions. This delayed reasoning about cause may help participants mitigate the risk of potential false discovery in visualization-only comprehension.
In some instances (31/120 R), participants formed 
initial interpretations from visualizations that hypothesized about causality,
yet explicitly noted how these hypotheses may be incorrect.
For example, after carefully examining the textual data for Farming and Underground Water (\autoref{fig:farmwater}), some found they might have focused too narrowly on the relation between farming and underground water in the chart, and corrected their inconclusive comprehension about the correlation between water levels and corn yields:}
    ``\emph{It seems like the chart was showing the results of dryland farming, rather than comparing the water decline to the results of the crops.}'' [$P24$].
\add{Such correction demonstrates that text can 
help resolve previous uncertainty or even false discovery from visual information: the participant conceptualized their observations as a sort of testable question or hypothesis to ``\emph{testing my last idea}'' [$P17$] and used the text to fill in the gaps that led to that thought.}

\fix{Further, we observed an \textbf{Aha!} moment---a moment of visible, positive knowledge construction and engagement---at least once for all thinkaloud participants when they encountered the text after the visualization.}
When the text explained an observation or hypothesis they had generated from the visual patterns, sometimes participants showed clear behavioral differences, such as heightened posture, responsive facial expression, and verbal exclamations---all of them (8/8 P) literally said ``\textbf{Aha!}'' at least once.
Due to space constraints, examples from in-person participants are available in \autoref{appd-inperson}.

When the visualizations show an obvious pattern like a change or outlier that can be easily captured in comparison, people are more likely to hypothesize a reason behind them.
In such cases, the text, especially when it provides externalized easy-to-understand reasons that do not appear in the charts' annotations, can help people make more sense of the data, lending context for deeper reasoning and preventing false discovery or conclusions.
However, when the chart is more complex, as in a distorted area chart (e.g., \autoref{fig:NYT7})
, or the additional contextual information is well-known (e.g., global warming, \autoref{fig:NYT11})
, these moments are less likely to arise: we found no evidence of any such ``Aha!'' behaviors in the two charts.

\vspace{0.2em}
\noindent \textbf{Summary}: Examining the differences between initial and revised comprehension, people seeing the \add{visualization-first} chain in multimodal communication 
exhibit \textbf{exploratory comprehension},
starting from discovery and data-oriented hypothesizing with the visualization and using the text to guide more precise and extended reasoning and inference.
People initially identified a diverse set of possible descriptions, keeping their language largely focused on the data and statistical patterns and expressing uncertainty when engaging in deeper reasoning. The text caused people to become more certain and integrate additional information, including causal reasoning and extending the described data to include additional patterns and observations. These extensions and movement towards rich, causal descriptions and ``Aha!'' moments of insight development reflect a hypothesis-driven process where people enable the data to lead their discovery process and use the text to assess and refine the resulting hypotheses. 

\subsection{\add{Text-First} Chain: From Narrative Framing to Visual Confirmation}
\label{sec-res-back}

In contrast to the \add{visualization-first} chain, participants who read the text first exhibited a top-down and narrative-driven reasoning process.
The text established a strong frame that guided precise and focused comprehension, but also constrained their subsequent interpretation of the visualization. Visualizations in the revised descriptions primarily served to confirm people's understanding of the data rather than offering new insight or observations.

\subsubsection{Initial Comprehension (Text Only): Establishing a Narrative Anchor}

When presented with only the text, participants' descriptions were summaries of the text's key claims and arguments (120/120).
Their initial description speaks to reasoning as processing a
predefined narrative rather than deeper engagement with the data.
This process created a strong mental anchor and a clear set of expectations for what the accompanying visualization would depict.

\begin{figure}[tbp]
    \centering
    \includegraphics[width=\linewidth]{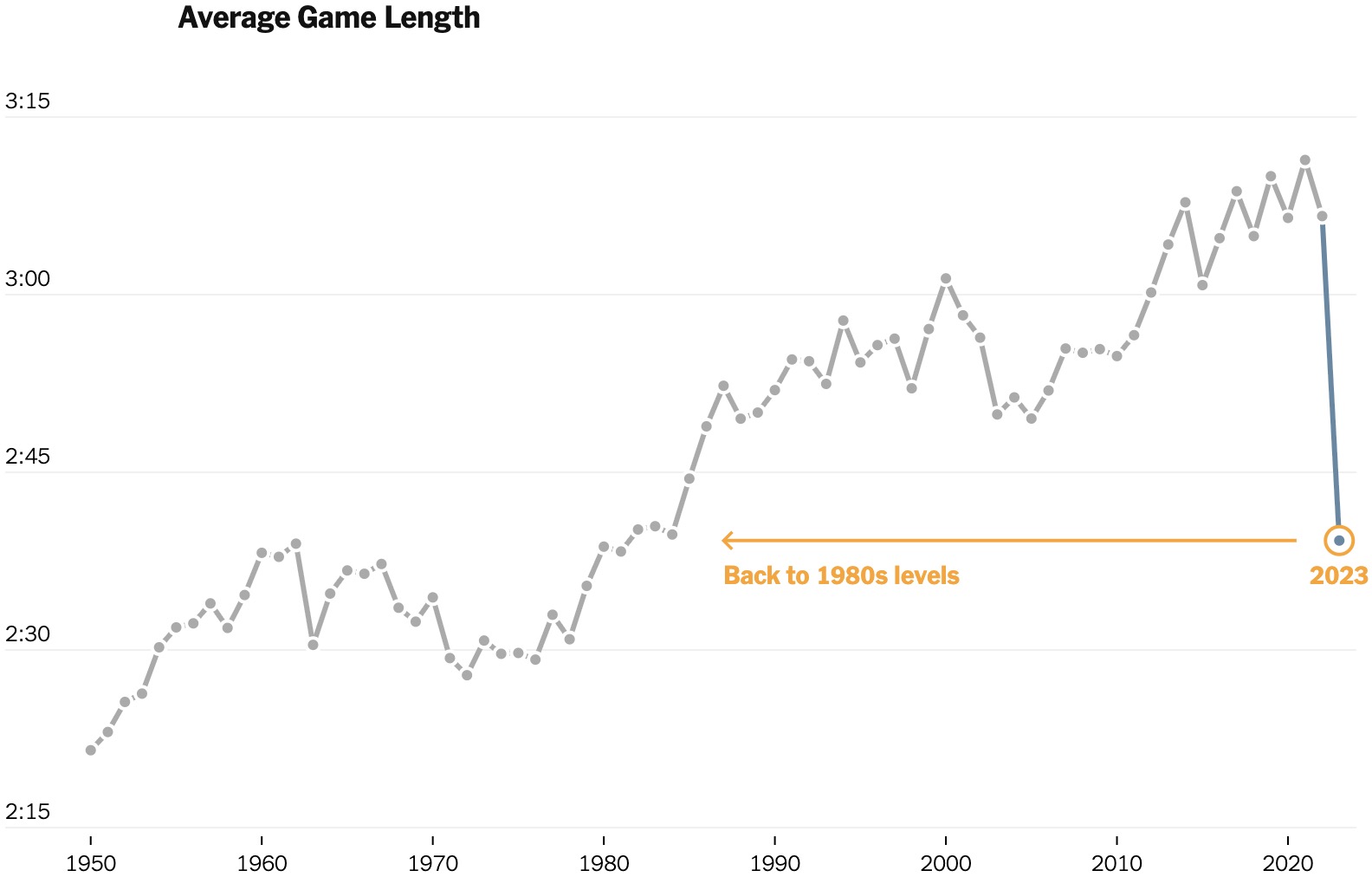}
\vspace{-1em}
    \caption{The visual graphic of Average Baseball Game Length~\cite{NYT2}.}
    \label{fig:baseball}
\end{figure}

As $P1$ wrote after reading paragraphs about Average Baseball Game Length (see \autoref{fig:baseball}), ``\emph{Baseball games are now as short as they were in the 1980s, due to rule changes.}''
All participants (10/10 P) exhibited little variation, \fix{solely focusing on describing the rule changes}, despite the volume of text presented.
For example, $P4$ noted ``\emph{Rule changes are switching baseball game times back to like they were in the 1980's.}''

\begin{figure}[t]
    \centering
    \includegraphics[width=\linewidth]{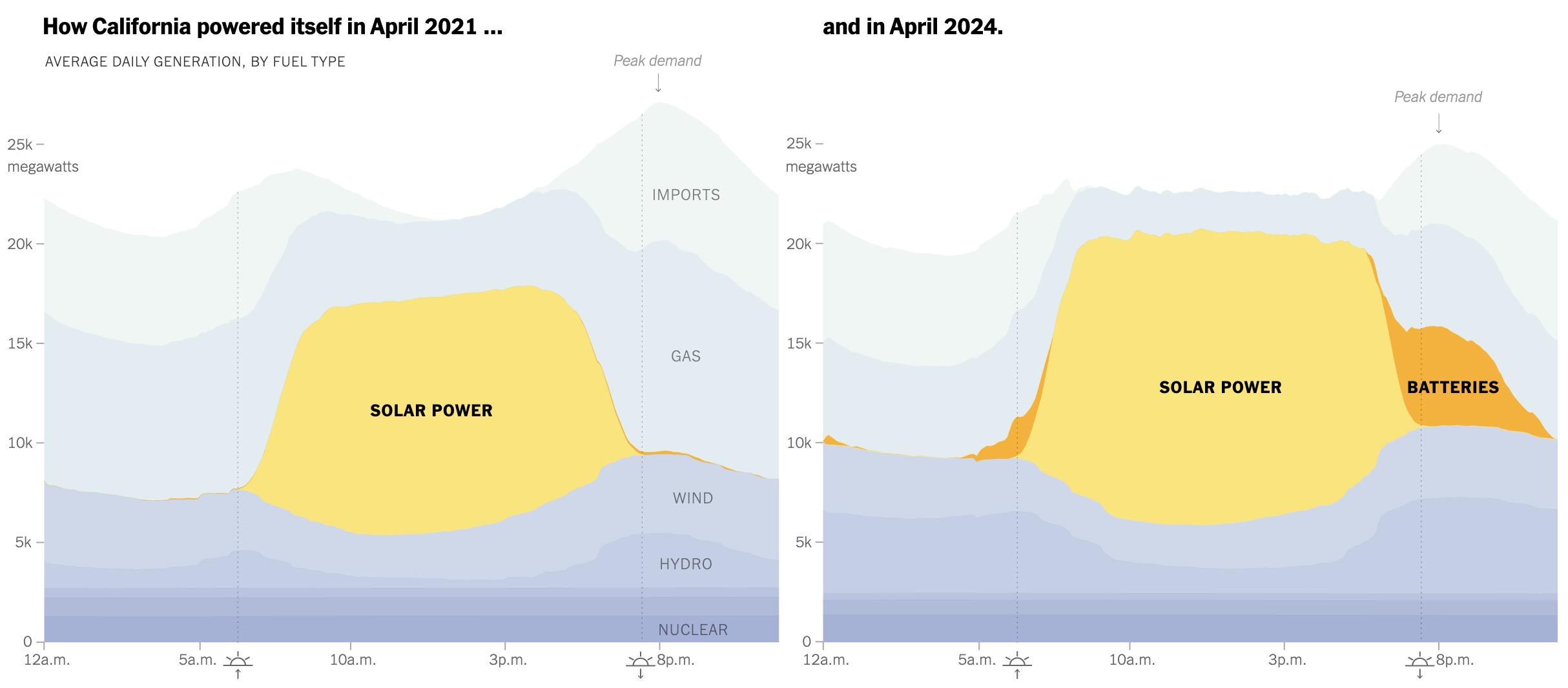}
\vspace{-1em}
    \caption{The visual graphic of California Electricity Power~\cite{NYT3}.}
    \label{fig:cal}
\vspace{-1em}
\end{figure}

Similarly, all descriptions \add{(10/10 P)} of California Electricity Power (see \autoref{fig:cal})
focused on the increasing construction of batteries to store California's solar power for usage at night. For example, 
    ``\emph{California built large batteries to store solar energy captured during the day so that people are still able to use...}'' [$P13$].

Unlike the visualization-first chain, where these two visualizations generated uncertain and varied hypotheses, in the text-first condition, 
people appear to engage in passive information reception, with the text serving as an anchor~\cite{tversky1974judgment}, building anticipation to see the ``\textit{proof}'' for the story they had just read in the visualization. This tended to manifest in the responses as a focus on high-level conceptual arguments rather than specific aspects or statistical patterns in the data.

\subsubsection{Revised Comprehension (Text + Visualization): Reinforcing the Predetermined Frame}

Upon seeing the visualization, people overwhelmingly tended 
to exhibit confirmatory behavior.
The chart was not viewed as a new source of information to be explored, but as evidence to validate the 
internalized narrative from the text.
Unlike in the \add{visualization-first} chain, where descriptions tended to expand and exhibit additional reasoning, adding additional information through the visualization in the \add{text-first} chain had little influence on participant comprehension despite offering more potential perspectives on the data.
\rev{While 40\% of the participants added contextual information such as visual patterns or values to better justify the claim, 60\% retained the same primary claim from their original comprehension.}
No descriptions challenged the underlying claims of the original text or revised their original understanding. 

Unlike the visualization-first chain, where many participants presented multiple hypotheses or patterns of interest, text-first participants tended to retain a singular focus in their responses (98/120 R). For example, in the Average Baseball Game Length stimulus (\autoref{fig:baseball}), all participants (10/10 P) 
retained their focus merely on the rules. $P1$ employed almost the same frame from their original comprehension noted above: 
    ``\emph{Baseball games are about as short as they were in the 1980s. Rule changes have shortened the games relative to where they were in the 1990s up until recently.}''.
Sometimes participants \add{(2/10 P)} added contextual information, like historically how the game length changes. For example, $P4$, who initially noted that the rules had brought game lengths closer to times from the 1980s, revised their comprehension with an observation about the overall historical trend from visualization, noting ``\emph{Baseball game lengths have historically been getting longer over time.}'' before the new rules.

We observed similar patterns in 
other examples. In California Electricity Power (\autoref{fig:cal}), all participants (10/10 P) again reported similar comprehension, \fix{focusing on batteries and time of day}. For example, $P13$ kept the main descriptions in their original comprehension described above: 
    ``\emph{The batteries California installed make it possible for people to continue using clean energy even after the sun goes down...}'' [$P13$].
$P15$ added context of how much population was impacted by this battery action, ``\emph{an additional 1/5 of the population are using stored energy from batteries...}''
but still focused on the same data patterns from their initial comprehension. 

Overall, the 
text 
acted as a form of perceptual filtering and framing: people tended to use visualizations as evidence of the core hypotheses from the text, but these representations led to little to no change to the overall description of the data.
Participants in this condition were significantly less likely to notice or comment on salient visual patterns---such as secondary trends, extremes, or outliers---if those features were not explicitly mentioned in the text, \fix{e.g., 69 visualization-first chain participants reported secondary patterns while only 22 in text-first chain reported secondary patterns}, exhibiting a sort of \emph{framing effect}~\cite{tversky1989rational}.
Their strategy was one of feature-matching, treating the visualization like a checklist to find evidence for the text's claims and integrating specific data from the visualization to refine the claims.
Consequently, the visualization was relegated to the role of an illustration, a graphical restatement of the text, rather than a tool for independent discovery and deeper inquiry.
As a result, people viewing multimodal data through a \add{text-first} chain were more likely to exhibit a \textbf{confirmatory comprehension} behavior, with knowledge being heavily constrained by 
the initial 
information in the text.

\fix{The behavior observed in the text-first chain during the thinkaloud participants additionally demonstrated this confirmatory approach.}
A common behavior, either stated explicitly or implied through their reasoning, among all participants was to use the visualization to simply say, ``\textbf{\emph{That (text) is correct.}}''
Please see \autoref{appd-inperson}
for more examples from in-person participants.

\vspace{0.2em}
\noindent \textbf{Summary}: \add{Participants' strategies in the text-first chain reflected 
a simple approach of using visualizations to find visual evidence supporting the story they had already been told.
This behavior 
indicates that the text-first chain 
induces
a strong framing effect leading to confirmatory comprehension, where the primary cognitive activity is matching visual patterns to a predefined textual narrative.}

\subsection{A Framework of Multimodal Data Comprehension}
\label{sec-res-behavior}

We can characterize the two chains of information in multimodal data comprehension using the 
observed differences among \add{four} features: cognitive process, primary activity, role of visualization, and outcome, as shown in \autoref{tab:comp}.
We elaborate on this characterization with 
quotes from the comprehension results and reasoning strategies employed in both chains, 
illustrating how the chains of information shape the process of multimodal comprehension.

\begin{table*}[htbp]
\caption{A Framework for Chains of Information in Multimodal Data Comprehension}
\vspace{-0.5em}
\label{tab:comp}
\centering
\begin{tabular}{|l|l|l|}
\toprule
\textbf{Feature}                       & \add{Visualization-First} Chain               & \add{Text-First} Chain             \\
\toprule
\textbf{Summary}     & Exploratory comprehension  & Confirmatory comprehension        \\
\hline
\textbf{Cognitive Process}     & Bottom-Up: From evidence to conclusion  & Top-Down: Framed understanding        \\
\hline
\textbf{Primary Activity}      & Hypothesis generation \& reasoning    & Narrative understanding \& reinforcement \\
\hline
\textbf{Role of Visualization} & A tool for discovery                    & An illustration for the text            \\
\hline
\textbf{Outcome}               & A synthesized multimodal understanding & A text-dominated scope-limited understanding \\
\bottomrule
\end{tabular}
\vspace{-1.5em}
\end{table*}

The way we present multimodal data changes how people engage with and develop knowledge through data. 
The \add{visualization-first} chain consistently fostered an exploratory and hypothesis-driven process as \textbf{exploratory comprehension}.
People acted like detectives, exhibiting a bottom-up reasoning process: forming initial hypotheses from visual evidence and then using the text to test and refine these hypotheses, gaining precision and integrating contextual information when presented with text (left part in \autoref{tab:comp}).
On the contrary, data comprehension from the \add{text-first} chain revealed a powerful framing effect to produce a top-down reasoning process: where the text established a narrative that people then sought to confirm with the visualization, changing their language but having little substantive impact on the information content of their comprehension (right part in \autoref{tab:comp}).
This \textbf{confirmatory comprehension} led to an evidence-matching process rather than open-ended exploration, with people seeking to ``fact-check'' the observations gained from text.
Please see \autoref{appd-operation}
for a more detailed discussion on operationalizing this framework.

\subsection{\add{Textual Cues within Visualizations}}
\label{sec-res-annotation}

\add{While our chain model divides text and visualizations into independent modalities, visualizations typically contain some amount of functional text 
(e.g., annotations, axis labels, titles, highlighted values) 
that can effect on people's understanding of data
as seen in prior work \cite{kong2018frames, xiong2022seeing, wang2024causal, stokes2023role, rahman2025exploring}.
Annotations specifically direct the reader's attention to particular values~\cite{rahman2025survey}, as seen in the Baseball example. We additionally examined the results from stimuli with and without annotations in the visualization-first chain to explore this affect.
Participants exposed to visualizations with clear textual cues in the visualization-first chain still engaged in a more exploratory process than those who read text first.
The textual cues acted as signposts within the data landscape, pointing out specific features, but they did not establish a full causal narrative or a rich contextual frame.
Participants did not universally include the annotated patterns in their responses, including the annotated patterns slightly more often in visualizations with annotations than without. 
For example, ``\emph{...temperature records...2021}'' was annotated in \autoref{fig:ustemp}, but more than half of participants (6/10) reported a different pattern or additional patterns. 
Some generated their own hypotheses about the annotated pattern, and some 
focused on other data patterns, such as ``\emph{1994 may be the coldest year}'' [$P16$], ``\emph{temperature records go back and forth}'' [$P3$], and ``\emph{cold temp records also appear in 2021}'' [$P9$].
The annotation in the Baseball Game stimuli (\autoref{fig:baseball}), noting where the numbers went ``\emph{back to 1980s levels}'' led to 6 of 10 participants reporting this same pattern in the visualization-first chain, while presenting the same information in the text of the text-first chain led all participants to report the pattern.}

That annotations did not fully constrain comprehension indicates a potential priming gradient: full textual narratives create the strongest framing effect, directing the reader's entire approach to the data.
Smaller textual cues, like visual annotations, provide a milder and more localized form of guidance, influencing which data features are attended to but not necessarily precluding broader exploration.
In our study, a clear visual annotation primed 40\%-60\% of participants to converge at a similar comprehension, while text primed nearly all readers towards a particular pattern in the corresponding visualizations.
Embedding textual cues in visualizations can enhance clarity without fully sacrificing the benefits of exploratory comprehension, whereas leading with a full paragraph of text can be a more powerful tool for guiding the audience toward a specific interpretation.

\section{Discussion}
\label{sec-discussion}

We examined how information chains in multimodal communication influence people's data comprehension. 
The simple act of presenting a visualization before or after its accompanying text creates two different cognitive behaviors.
Presenting the visualization first fosters a bottom-up and \textbf{exploratory} sensemaking process, where readers form hypotheses from visual evidence.
In contrast, presenting the text first promotes a top-down \textbf{confirmatory} process in which a powerful framing effect guides and constrains the reader's interpretation of the visualized data.

\subsection{Theoretical Implications}

Our work provides empirical evidence connecting long-standing theories in cognitive science and guidelines of narrative visualizations within the context of multimodal data communication. 
The behavior exhibited in the \add{text-first} chain 
reflects \emph{priming}~\cite{lashley1951problem} and \emph{framing effects}~\cite{tversky1989rational}.
The initial text acts as a strong prime \fix{(to strongly convey a piece of information related to the data)}, establishing a narrative frame \fix{(reader understands this piece as the underlying message before seeing charts)} that anchors the reader's expectations.

This framing effect has downstream consequences for how people make sense of the presented information. 
For example, people may become subject to 
potential \emph{confirmation bias}~\cite{wason1960failure}, seeking out visual evidence that supports the predefined 
narrative while overlooking other information, especially information that may be ambiguous, unrelated, or even contradictory. 
They may overemphasize patterns associated with the text, as in the \emph{Curse of Knowledge} \cite{xiong2019curse} or even override or falsely interpret contradictory observations in the data, as with misleading titles \cite{kong2018frames} or strong beliefs about presented variables~\cite{wang2026contextualization}. This framing effect also limits engagement and agency: people limit their interaction with the data to confirming the claims provided in the text at the cost of exploring their own, more diverse impressions of the data. 

In contrast, the \add{visualization-first} chain tends to promote greater agency, reducing framing effects at the cost of potentially greater upfront uncertainty. Despite being designed to communicate points aligned with the text descriptions, participants expressed a wider array of statistical observations in the \add{visualization-first} chain. Their strategies reflect this discovery-first process as well, noting the freedom and flexibility the visualization allows when presented before the accompanying text. This result confirms recent observations that the conclusions we draw from data may vary, even for relatively simple or familiar visualizations \cite{quadri2024do, bearfield2024same, williams2023data}. However, this agency comes at the cost of increased cognitive effort on the part of the reader and potential misalignment between the objectives of the chart and what readers learn from it.  
\add{Further, visualization-first chains can be regarded as a simpler everyday version of visual data exploration.
The visualization acts as a lightweight exploratory interface: people first engage in pattern-seeking and hypothesis generation from the visual evidence alone, a classic exploratory activity~\cite{battle2023we, gotz2008characterizing, battle2019characterizing}.
The subsequent text then functions as a key step in the sensemaking loop, providing the context and causality needed to validate, refine, or correct those data-driven hypotheses, like guidance in visual data exploration~\cite{wang2024beyond, narechania2024designing, ceneda2016characterizing, stokes2024give}.}

Viewed collectively, our findings add a critical layer to data comprehension research, which has traditionally focused on interpreting charts in isolation.
Adding additional context can further shape the information that people extract from visualizations, whether as part of exploratory knowledge formation in the \add{visualization-first} chain or confirmatory behaviors in the \add{text-first} chain. 
The insights people extract from a visualization are not static but highly malleable and subject to the influence of preceding information~\cite{battle2023we}. This malleability emphasizes the importance of assessing data interpretation across modalities, echoing Hearst's call for a co-equal treatment of text and graphics \cite{hearst2023show}. 

Furthermore, these findings provide a mechanistic explanation for the widely-applied narrative structures in visual data storytelling identified by Segel \& Heer~\cite{segel2010narrative}.
The \emph{author-driven} narrative paths, where a story is explicitly laid out for the reader, function similarly to our \add{text-first} chain by establishing a controlled frame that the visuals then serve to illustrate.
Conversely, the initial exploratory phase of our \add{visualization-first} chain reflects a likely \emph{reader-driven} exploratory path they describe, allowing people to first distribute attention on their own to comprehend visualized data, then the corresponding insights are refined through engagement with the complementary text modality.
We show that these information chains form a primary tool authors can use to shift between these narrative modes, either guiding readers toward a specific conclusion or inviting them into open-ended discovery.

\subsection{Guidelines Across Data Communication Goals}

The differences between exploratory comprehension and confirmatory comprehension, observed in the two chains of information, have important practical implications for creating or consuming data-driven content.
Our findings are not prescriptive but rather descriptive, urging creators to be intentional about their design choices.

    To \textbf{encourage exploration, discovery, and agency}, %
    consider designing communications that lead with the visualization. For example, one might put the chart at the top of an article, use it as the opening slide or view in a slideshow or scrollytelling story, or leverage visual salience to strongly pull the reader's attention to the graphic.
    This approach is ideal for any use cases that need people to first engage deeply with the data, such as guiding readers through a range of 
    analytical questions, open-ended data features, or situations where the goal is to empower the audience to draw their own conclusions before being presented with a narrative.
    Designing for the \add{visualization-first} chain conceptually invites readers into a process of inquiry and subsequent dialog with the author through the ensuing text.
    
    To \textbf{communicate a specific and intentional message} and ensure it is received as intended, 
    consider leading with text. 
    Text-first leverages the power of framing to focus people's attention on what matters most in the data. This approach may be useful in cases where a communication is intended for persuasion or for communicating complex concepts, such as explaining scientific phenomena to a lay audience. 
    However, this power comes with significant ethical responsibility.
    Creators must be aware that this choice minimizes the likelihood of alternative interpretations and should ensure their frame is a faithful representation of the data to avoid ethical issues, such as perpetuating incorrect or biased information
    \cite{roozenbeek2020susceptibility}.
    \add{Alternatively, they may choose to use visual annotations to reduce the priming power while still offering active direction towards critical elements of the data.}

The results also indicate a need for data literacy curricula %
to better address multimodal contexts.
    As a complement to assessing and teaching how effectively people can read a chart in isolation, 
approaches can teach how to recognize and critically assess the influence of text accompanying a chart. 
Developing proactive awareness may encourage deeper critical reflection in the \add{text-first} chain and potentially help mitigate the reduced analytical agency engendered by such designs in the future.

\subsection{\fix{Format vs. Amount of Information in Real-World Design (In)Coherence Across Modalities}}

\add{Our stimuli used a typical real-world multimodal data format (news articles) instead of fully-controlled visualizations designed for research purposes.
A critical consideration from this approach 
is the inherent division of labor in authoring news media, where articles are typically developed by small teams. Journalists typically author the text and visual designers or graphics editors create the accompanying charts.
This separation of roles could have direct implications for the coherence between the modalities.
}

\add{The text used in our study, while directly related to the visuals, was typically authored to function as a coherent part of the broader article.
Unlike alternative texts or captions that only describe visualizations~\cite{lundgard2021accessible, zong2022rich}, they provide narrative context, contextualized explanations, and real-world significance that extend beyond the visualizations' immediate purview.
In contrast, the visualizations may be designed to illustrate and support the specific narratives presented in the text. 
The visualizations we employ are, by nature, not typically intended to be fully stand-alone artifacts but rather
potent illustrations within a specific textual context.}
\rev{Therefore, the text fragments we tested may not be representative of the full breadth of statistical details present in the text.}

This asymmetry in design intent is a common case of real-world multimodal communication---the interaction between text and visualization aims to enhance communication and data comprehension \cite{hearst2023show}. 
In the text-first chain, this coherence works as intended: the text establishes a narrative frame tailored for the visualization to then illustrate and concretize.
This therefore leads to the powerful confirmatory comprehension we observed, as people only focus on part of the visualization that fits into the space prepared for it by the text.

However, in the visualization-first chain, we see alternative comprehensions arising from 
a visualization, even though the visualization was designed for a specific narrative.
Participants lacked the scaffolding of the narrative framing that the article
intended to convey.
This, instead, allowed them engage in an exploratory comprehension process, generating their own diverse hypotheses and patterns, some of which aligned with the article's intended narrative and others which did not, resulting in a broader range of insight and greater reliance on their own agency to make sense of the data.

It is worth noting that comprehension may also be shaped by the underlying design intent and the degree to which each modality is optimized to function independently versus as an integrated unit.
Some data stories intend the visualization to carry the weight of the narrative.
Further, visualizations tend to carry more statistical information than text within the same amount of visual space.

\rev{This asymmetry accurately reflects real-world data journalism: text typically offers a more focused message, but lacks context. However, this also introduces an inherent confound between the delivery modality and the amount of information provided.
Future work should decouple these factors by testing conditions with absolute information parity---for instance, comparing visualization chains against an alternative text condition that systematically enumerates every data trend in pure prose.}

\subsection{\fix{Limitations and Future Directions}}

\rev{Our work explores 
multimodal data comprehension of visual and textual information. 
While this experiment offers preliminary insight into how chains of information can influence data interpretation, it also raises new opportunities to understand how alternative approaches to multimodal communication can shape data comprehension. }

\subsubsection{\rev{Beyond Sequential Viewing}}
\rev{While our sequential model divides presentation order into two simple opposing chains, real-world visualization consumption may not behave as cleanly. 
Beyond linear sequential viewing, people can dynamically alternate between visuals and text when both modalities are presented simultaneously, 
actively moving their attention back and forth between a textual passage and an accompanying data visualization to incrementally piece together a unified understanding. However, attention is inherently a sequential process, and we typically see visualizations in the media designed to support consumption of either text first or images first.
Future research should explore how alternating attention influences interpretation and how different designs might encourage different attentional behaviors. }

\rev{Based on our findings, we hypothesize that this alternation behavior may manifest as a continuous loop of cross-modal verification.
When a reader's primary attention is anchored in the textual narrative, they treat the visualization as a localized reference point to locate the corresponding visual feature from the text.
Conversely, a reader might use the visualization as an exploratory map, identifying an unexpected spike, dip, or cluster, and then scan the nearby text to find an explanation or causal mechanism that accounts for that visual anomaly, especially if an image is used as a teaser or made particularly salient by design.}

\rev{While our chains of information model treats text and visualizations as sequential blocks of information, real-world media often relies on charts that tightly blend these modalities dynamically, such as data comics and interactive scrollytelling.
These designs may force people to rapidly switch focus between reading words and decoding charts.
Future research should investigate how sequential viewing behavior operates in these hybridized formats.}

\subsubsection{LLM-Lead Information Delivery}
\fix{Real-world information chains are increasingly being reshaped by assistive technologies
such as LLMs, which frequently generate textual summaries or interpretations before a user interacts with the original visual graphics.
This shift potentially places users more commonly in a text-first chain, which our study shows leads to a 
using visualizations for confirmatory rather than exploratory behavior. }

\fix{Future research should empirically measure the trade-offs between the efficiency of LLM-generated summaries and the cognitive biases they may introduce. Text-first chains established a narrative anchor 
that led people to engage in a confirmatory behavior
rather than open-ended discovery.
In the context of LLMs, this framing effect could significantly amplify susceptibility to AI hallucinations and loss of human analytical agency, raising the importance of understanding how the order of information delivery 
when designing exploratory AI assistants.}

\subsubsection{\rev{Graph Literacy, Accessibility, and Complexity}}
\rev{Our crowdsourced sample represents a sighted audience with general literacy level.
Individuals with lower graph literacy may be more vulnerable to the strong framing effects observed in the text-first chain~\cite{chen2024unmasking}.
While we found no notable differences across self-reported visualization familiarity, people with substantive experience working with data might approach data differently.
For blind or low vision (BLV) individuals, the way they perceive, process, and engage with visual information primarily using alternative text~\cite{zong2022rich} might lead to personalized behaviors and frictions combining visual and textual information~\cite{elavsky2025towards}.}

\rev{Our study focused on information-rich communicative graphics.
When a visualization is inherently rich and complex, it demands higher cognitive effort, which may naturally trigger a state of exploratory behavior in the visualization-first chain~\cite{quadri2024do, bearfield2024same}.
However, this dynamic might look different for very simple bar charts or single-line graphs whose entire message can be gleaned in a brief glance.
For trivial charts, two chains may collapse into a uniform comprehension.
Future work should investigate the relation between complexity and comprehension behavior.}

\section{Conclusion}
\label{sec-conclusion}

Real-world data communication frequently combines visualizations and text. 
Our research demonstrates that the interplay between these two modalities 
is profoundly influenced by a simple variable: which one speaks first.
By introducing and validating the paradigm of \emph{chains of information in multimodal data comprehension}, we provide empirical evidence of how the composition of multimodal communication influences how people engage with data. Seeing a visualization first fosters exploration, while reading the text first triggers confirmation.
Understanding and deliberately harnessing the power of this sequence is not merely a design choice: 
it can help designers create more effective, ethical, and insightful data communication.

\acknowledgments{
We thank the reviewers for their insightful comments.
This work was supported by the National Science Foundation under grant NSF IIS-2046725 and NSF IIS-1764089.
}

\bibliographystyle{abbrv-doi-hyperref-narrow}

\bibliography{Sections/main}

\appendix
\section*{Appendix}

\section{Stimuli Selection}
\label{appd-stimuli}

\autoref{tab:chartall} and \autoref{tab:topic} show overviews of all reviewed visual graphics and selected topics from the final stimuli.
Below shows the stimuli selection criteria and process.

\begin{itemize}
    \item \add{\textbf{Common and Familiar}: We focused on common visualization types and topics. We removed those that may be less familiar (such as rotated parallel coordinates) or too complex (such as a combination of area charts and connected scatterplots) for the general public. We avoid data on topics that need specific domain knowledge (e.g., aerospace technology) or may elicit political sentiment (e.g., election data). 53 visualizations were excluded under this criterion.}
    
    \item \add{\textbf{Clarity and Readability}: Charts were required to be legible at the study display size (e.g., readable axis labels and legends at 16 or 14 pixel minimum respectively). 30 visualizations were excluded under this criterion.}

    \item \add{\textbf{Stable Static Representation}: Only static images (no interactive versions, such as animated flow diagrams and Sankey charts) were used to avoid potential confounds.
    Any charts that heavily relied on interaction for core content were excluded. 25 visualizations were excluded under this criterion.}

    \item \add{\textbf{Textual Specificity}: The accompanying paragraphs needed to include sufficient explanatory or contextual information (not merely a one-line caption) so that text could plausibly change or refine a reader's comprehension. Charts with purely decorative or unrelated text were excluded. 11 visualizations were excluded under this criterion.}
    
    \item \add{\textbf{Neutral Attribution}: To avoid bias from strong source signals, articles and graphics that explicitly named potentially polarizing, professional, or authoritative organizations or individuals in proximate text were avoided as these organizations may shift how readers weigh different pieces of information (e.g., weighing explicit expert interpretation more heavily than raw data). 3 visualizations were excluded
    For the rest, identifying information was removed when present to mitigate potential bias.}

    \item \add{\textbf{Topic and Visualization Diversity}: We ensured a range of diverse topics and visualization types (as summarized in \autoref{tab:chart} and \autoref{tab:topic}) to avoid results driven by a single domain or visualization type. Stimuli were randomly sampled from visualizations meeting the above criteria, with 
    the final selection limited to no more than three stimuli for each visualization type and up to two stimuli on a same topic.}

\end{itemize}

\begin{table*}[htbp]
\renewcommand{\arraystretch}{1.2}
\centering
\caption{\add{Visualization types and counts for NYT visual graphics published from 2020 to 2024. The \emph{Others} chart type includes 
non-traditional charts, such as treemaps, pictograms, parallel coordinates, flow diagrams, and Sankey diagrams, that seldom appeared in the corpus.}}
\label{tab:chartall}
\begin{tabular}{|c|c|c|c|c|c|c|}
\hline
Visualization Types & Map & Line Chart & Bar Chart  & Area Chart & Scatterplot & Others \\
\hline
Counts     & 110    & 22   & 21     & 13  & 6  & 9 \\
\hline
\end{tabular}
\end{table*}

\begin{table*}[htbp]
\renewcommand{\arraystretch}{1.2}
\centering
\caption{Topics and counts in the chosen visual graphics.
}
\label{tab:topic}
\begin{tabular}{|c|c|c|c|c|c|c|c|c|c|c|}
\hline
Topic & Temperature & Energy & Farming & Health & Sports & Friends & \add{Daily Activities} & Diet & Disaster & Snowfall \\
\hline
Counts & 2   & 2     & 1  & 1  & 1 & 1 & 1 & 1 & 1 & 1 \\
\hline
\end{tabular}
\end{table*}

\section{Insights from In-Person Participants}
\label{appd-inperson}

\subsection{Visualization First Chain -- The Aha! Moment}

We observed an \textbf{Aha!} moment---a moment of visible, positive knowledge construction and engagement---at least once for all thinkaloud participants when they encountered the text after the visualization.
When the text explained an observation or hypothesis they had generated from the visual patterns, sometimes participants showed clear behavioral differences, such as heightened posture, responsive facial expression, and verbal exclamations---all of them (8/8 P) literally said ``\textbf{Aha!}'' at least once.

\add{For instance, after studying the visualization on California Electricity Power (\autoref{fig:cal}), all participants (4/4 P) started to wonder why ``\emph{batteries}'' increase in usage in the second half of the chart, which the text in the second phase of the chain explained (``\emph{soak up excess solar power during the day and store it for use when it gets dark.}'').
For example, $I3$ hypothesized aloud about batteries ``\emph{saving solar power}'' as the cause. 
$I8$ noticed the X-axis of time and the ``\emph{more flatten power usage during the day}'', assuming 
the corresponding increase in battery usage was due to ``\emph{getting over-generated daytime power}''.
Upon reading the text, they noted feeling ``\emph{surprised}'' or ``\emph{so smart}'', and both praised themselves for 
correctly hypothesizing the cause. 
Two other participants 
focused on other reasons for the rise like ``\emph{transferring gas usage}.''
After seeing the text, they responded 
verbally in ways indicating their new understanding, such as ``\emph{Aha, that makes sense}'' [$I1$] or ``\emph{So that explains this big change}'' [$I5$].
Such moments of validation highlighted 
how people were able to pair visual evidence with text to support hypothesis generation and validation in the visualization-first chain. 
}

\add{These moments not only appear in confirming or explaining a hypothesis, but also when reducing confusion in uncertain understanding and 
correcting potential false discoveries.
In the Average Baseball Game Length stimulus (\autoref{fig:baseball}),
the text 
explained the annotated 
reduction in game length in 2023
is because of ``\emph{a series of rule changes}'' [text].
All participants first noted the overall increasing trend in the historical data but then 
expressed confusion in the sudden reduction
and uncertainty in their comprehension, noting feeling ``\emph{strange}'' [$I2$] or ``\emph{not confident}''  [$I3$] about the apparent anomaly.
After seeing the text that \fix{explained rule changes at that year}, participants felt they better understood the data, noting ``\emph{Oh I totally missed it}'' [$I2$] and that the change now ``\emph{definitely makes sense}'' [$I6$].
These instances illustrate a different but equally significant ``Aha'' moment: one of correction.
The exploratory path is both about forming correct hypotheses
and using the text 
as a corrective mechanism, resolving the ambiguity of an isolated visualization and helping to fill in complexities or uncertainties in data which are difficult to resolve through visualization alone~\cite{hearst2023show}.}

When the visualizations show an obvious pattern like a change or outlier that can be easily captured in comparison, people are more likely to hypothesize a reason behind them.
In such cases, the text, especially when it provides externalized easy-to-understand reasons that do not appear in the charts' annotations, can help people make more sense of the data, lending context for deeper reasoning and preventing false discovery or conclusions.
However, when the chart is more complex, as in a distorted area chart (see \autoref{fig:NYT7}), or the additional contextual information is well-known (e.g., global warming, \autoref{fig:NYT11}), these moments are less likely to arise: we found no evidence of any such ``Aha!'' behaviors in the two charts.

\subsection{Text First Chain -- That (Text) is Correct}

\fix{The behavior observed in the text-first chain during the thinkaloud participants additionally demonstrated this confirmatory approach.}
A common behavior, either stated explicitly or implied through their reasoning, among all participants was to use the visualization to simply say, ``\textbf{\emph{That (text) is correct.}}''

Let's review the same examples that other participants see visualization-first feel ``\emph{Aha}''.
\add{In California Electricity Power (\autoref{fig:cal}), upon viewing the chart after reading the text, all participants simply noted, ``\emph{That is correct}'' [$I2$] or ``\emph{Batteries do help}'' [$I7$].
Unlike the visualization-first chain's \textit{Aha} moments, none of them reported findings like how the gas usage was reduced or how the daily total power usage was flattened. %
Similarly, in Average Baseball Game Length (\autoref{fig:baseball}), instead of analyzing the increasing trend from the zero point, which was a critical piece of the ``Aha!'' moments in the visualization-first chain, they immediately viewed the data points in 2023 and stated like, ``\emph{the new rules make a big change}'' [$I1$] or ``\emph{this [the downward shift in value] is what happened in 2023 turned baseball back to 1980}'' [$I5$].
Overall, the visualization was not treated as a source of new insight but as a source to validate the primed knowledge.
We did not observe any ``Aha!'' moments when viewing the visualizations in the text-first chain. In contrast to the visualization-first chain, the text-first thinkaloud responses
to the visualizations did not provide moments of new discovery or synthesis or expressions of 
surprise or validation, but instead universally expressed alignment with the original text.}

\section{Operationalizing Multimodel Data Comprehension Framework}
\label{appd-operation}

In this appendix, we introduced a more detailed discussion on how the proposed multimodel data comprehension framework in \autoref{sec-res-behavior} can characterize user behaviors in the two chains.

\subsection{Characterizing the \add{Visualization-First} Chain}

As summarized in the left half of \autoref{tab:comp}, the \add{visualization-first} chain induced a bottom-up style reasoning process, where people typically started with raw visual evidence and worked their way toward a conclusion about the data through text, usually associated with causation.
The visualization acts as a tool for discovery, capable of revealing insights without restrictions.
People's experiences and subsequent descriptions reflected a progression from curiosity and inquiry to resolution.
Overall, the \add{visualization-first} chain led to a synthesized understanding where evidence from both modalities were woven together.
The 
final comprehension tended to be a rich and balanced combination of visual patterns and narrative context.

\begin{figure}[htbp]
    \centering
    \includegraphics[width=\linewidth]{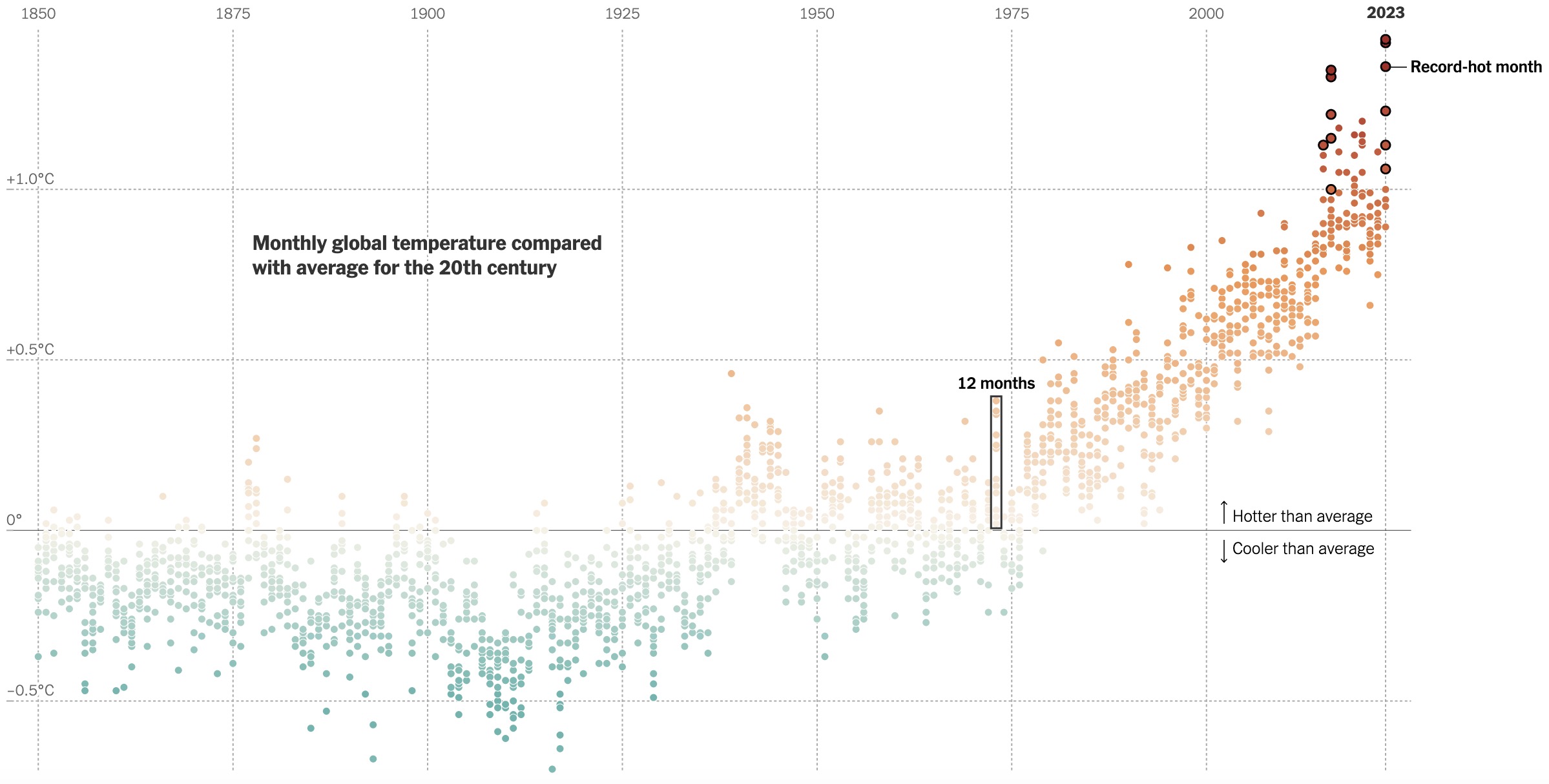}
    \caption{The visual graphic of Global Temperature.}
    \label{fig:global}
\end{figure}

For example, people who first saw a visualization of the Global Temperature (see \autoref{fig:global}), reported a diverse range of initial comprehensions that heavily relied on different visual patterns, such as: 
\begin{quote}
    ``\emph{Starting around 1975, monthly global temperatures have exceeded the average for the 20th century.}'' [$P57$], 
    
    ``\emph{Monthly global average temperatures have increased over 1 degrees celsius compared to the average for the 20th century. Almost all of the rise has been since the early 70's when a steady rise in average temperatures began with the most recent year showing 5 months at record high levels.}'' [$P59$].
\end{quote}
Initial comprehensions for this chart varied significantly, including
extremes, trends, changes, value ranges, and periodic patterns.
After viewing the accompanying text, participants integrated the textual information to complement their understanding of visual patterns. The two comprehension stated above were revised to: 
\begin{quote}
    ``\emph{Global temperatures have increased significantly in the last 150 years, with monthly records being set consistently in the recent past.}'' [$P57$], 
    
    ``\emph{2023 was the hottest year on record since record keeping began 174 years ago. 2023 was the hottest ever record and the final 5 months were the hottest months on record.  Global average temperature in 2023 was nearly 1.5 celsius higher than the average for the 2nd half of the 19th century and warmer than the 2nd hottest year in 2016 by a large margin.}'' [$P59$].
\end{quote}
Their revised comprehension includes additional information, like ``\emph{in the last 150 years}'' and ``\emph{the hottest year since 174 years ago}'' to better contextualize the visual patterns, but still covers a range of different observations in visual patterns, including patterns that were not discussed in the text.

People's self-reported  \textbf{reasoning strategies} 
reflected this overarching 
process in the \add{visualization-first} chain.
For example, $P43$ 
noted in the Disaster and Climate Risks example (see \autoref{fig:NYT12}) that 
\begin{quote}
    ``\emph{When I see the chart first, I get an immediate visual understanding of the geographic distribution and intensity of climate risks. It shows where different threats like wildfires, water stress, and hurricanes are concentrated. Then, when I read the text, it helps me understand the broader context, explains the causes and impacts, and connects the data to real-life consequences, making the information more meaningful.}''
\end{quote}
For the US Electronic Generation example (see \autoref{fig:NYT4}), $P7$ described that 
\begin{quote}
    ``\emph{Looking at the chart, I first see coal decline and the rise of natural gas and wind. The text then explains when and why this shift happened, showing how it reduced emissions and pollution. Together, they make the trend and its impact much clearer.}''
\end{quote}
These %
strategies match the exploratory comprehension pattern described above: participants first form a preliminary understanding of the overall data from visualized patterns, then the text provides context and explanations of the causes behind data patterns to help people better comprehend the data and expand their awareness and understanding.

\subsection{Characterizing the \add{Text-First} Chain}

Conversely, illustrated in the right half of \autoref{tab:comp}, the \add{text-first} chain is more likely to produce a top-down reasoning process.
People began with the abstract claims from the text and used the visualization to find concrete evidence for that predefined claim.
The resulting comprehensions suggest participants actively trying to validate the text's narrative, and are often revised with a sense of confirmation.
The visualization, therefore, was treated as an auxiliary illustration of the textual information.
People's experience was less about exploring something new and more about reinforcing what was already learned from the text.
This produced a confirmatory understanding, where the textual narrative dominated 
people's perception of the data and potentially masked other details in the visualization, similar to the Curse of Knowledge \cite{xiong2019curse}.
The final comprehension was typically a restatement of the text's main point.

Viewing the same Global Temperature stimulus described above (\autoref{fig:global}), people's comprehensions 
in the \add{text-first} chain focused on the hottest year and the impact of climate change, in contrast to the statistical orientation seen in the \add{visualization-first} chain.
For example, 
\begin{quote}
    \add{``}\emph{2023 was the hottest year ever in at least 174 years, maybe even longer. Every month from June to December set new heat records. People are worried that global warming is happening faster than before.}'' [$P42$],
    
    \add{``}\emph{Climate change is having an effect on our environment through the documented change in temperature which lends to the idea of global warming. 2023 was the warmest year in the past 174 years when temperatures in June surpassed previous records.}'' [$P48$].
\end{quote}
After viewing visualizations, they gathered evidence to support the claims from in their initial comprehension and closing with confirmatory language explicitly framing the added statistical information as evidence: 
\begin{quote}
    ``\emph{The chart shows that the Earth is getting warmer over time. Since the 1970s, there have been more hot months than cool ones. In 2023, the hottest month ever was recorded, showing that global warming is happening.}'' [$P42$],
    
    ``\emph{Not only was June of 2023 the hottest month on record in the last 174 years but temperature average has gradually increased during a 12-month span since 1850 which lends to further evidence of global warming due to climate change.}'' [$P48$].
\end{quote} 
Their final comprehensions heavily focused on claims about climate change, which is a major piece of information from the text but not explicitly described in the chart.

Similarly, people reported reflective insights as core to their \textbf{reason strategy} for the data from the \add{text-first} chain.
For example, $P25$ said, 
\begin{quote}
    ``\emph{I like reading an explanation in text first, then having it enhanced with charts. I like the text to tell me what I'm going to see and to help me interpret what the chart means. With text first, I grasp charts much more quickly.}''
\end{quote}
$P45$ stated, 
\begin{quote}
    ``\emph{When looking at the text first I am gathering the biggest takeaway and then I look at the chart and try to see how I can prove that takeaway with the data.}''
\end{quote}
$P58$ simply concluded, ``\emph{I draw a conclusion and see if the chart confirms it.}''
All of these strategies reflected a framing or confirmation-based behavior, where the understanding of the chart serves as a secondary supplement to the text.

\subsection{Framework Summary}

Overall, applying our framework to the data comprehension and self-reported strategies employed by participants demonstrates that multimodal data comprehension can lead to different comprehension patterns depending on the order in which people attend to each modality. 
The \add{visualization-first} chain promotes an exploratory bottom-up behavior to allow exploratory comprehension, enabling people to actively explore visual evidence, form independent hypotheses, and seek textual context for resolution.
Conversely, the \add{text-first} chain elicits a top-down and confirmatory behavior, where the initial text establishes a strong frame of narrative, causing people 
to primarily seek to validate that frame by matching it to the visual evidence.
This difference determines whether a visualization is treated as a tool for discovery or as an illustration of a story already told.

As summarized in the left half of \autoref{tab:comp}, the \add{visualization-first} chain induced a bottom-up style reasoning process, where people typically started with raw visual evidence and worked their way toward a conclusion about the data through text, usually associated with causation.
The visualization acts as a tool for discovery, capable of revealing insights without restrictions.
People's experiences and subsequent descriptions reflected a progression from curiosity and inquiry to resolution.
Overall, the \add{visualization-first} chain led to a synthesized understanding where evidence from both modalities were woven together.
The 
final comprehension tended to be a rich and balanced combination of visual patterns and narrative context.

Conversely, illustrated in the right half of \autoref{tab:comp}, the \add{text-first} chain is more likely to produce a top-down reasoning process.
People began with the abstract claims from the text and used the visualization to find concrete evidence for that predefined claim.
The resulting comprehensions suggest participants actively trying to validate the text's narrative, and are often revised with a sense of confirmation.
The visualization, therefore, was treated as an auxiliary illustration of the textual information.
People's experience was less about exploring something new and more about reinforcing what was already learned from the text.
This produced a confirmatory understanding, where the textual narrative dominated 
people's perception of the data and potentially masked other details in the visualization, similar to the Curse of Knowledge \cite{xiong2019curse}.
The final comprehension was typically a restatement of the text's main point.

\section{Codebook}

\label{appd-code}

\autoref{tab:code} shows the employed codebook and coded examples with users' quotes.
Our codebook consisted of two groups of codes: \textit{Comprehension-Related Codes} (e.g., \textit{INITIAL\_VIS\_COMPREHENSION} codes how people's comprehension based on the visualization first reflects the data, and \textit{INFO\_GAIN\_FROM\_TEXT} codes new information that the revised comprehension gained from the text)
and \textit{User Experience and Strategy-Related Codes} (e.g., \textit{STRATEGY\_VIS\_FIRST} codes user described a specific strategy for interpreting the \add{visualization-first} chain, and \textit{PREFERENCE\_VIS\_FIRST} codes user explicitly states a preference for seeing the visualization first).

\noindent 1. \textbf{Comprehension-Related Codes}:
\begin{itemize}
    \item INITIAL\_VIS\_COMPREHENSION: Initial comprehension based on the visualization first.
    \item INITIAL\_TEXT\_COMPREHENSION: Initial comprehension based on text first.
    \item INFO\_GAIN\_FROM\_TEXT: The revised comprehension shows significant new information gained from the text (e.g., specific numbers, context, ``\textit{why}'').
    \item INFO\_GAIN\_FROM\_VIS: The revised comprehension shows new information gained from the visualization (e.g., visual patterns, magnitude of change).
    \item CONFIRMATION: The second modality is used to confirm what was learned from the first.
    \item CORRECTION: The second modality corrects a misinterpretation from the first.
    \item SYNTHESIS: The participant integrates both visual and textual information into a balanced coherent final understanding.
\end{itemize}

\noindent 2. \textbf{User Experience and Strategy-Related Codes}:
\begin{itemize}
    \item PREFERENCE\_TEXT\_FIRST: Participant explicitly states a preference for seeing the text first.
    \item PREFERENCE\_VIS\_FIRST: Participant explicitly states a preference for seeing the visualization first.
    \item DIFFERENCE\_VIS\_VS\_TEXT: Participant provides commentary on the perceived difference in the two approaches.
    \item STRATEGY\_VIS\_FIRST: Participant describes a specific strategy for interpreting a chart without context (e.g., ``\textit{I look for a trend}'').
    \item STRATEGY\_TEXT\_FIRST: Participant describes a specific strategy for interpreting a chart with prior context (e.g., ``\textit{I look for the numbers mentioned in the text}'').
\end{itemize}

\begin{table*}[htbp]
    \centering
    \caption{\add{Codebook used to analyze the data, themes, and behaviors in our analysis user quotes examples.}}
    \label{tab:code}
    \begin{subtable}{\textwidth}
        \centering
        \caption{Comprehension-Related Codes and User Quotes}
        \begin{tabular}{>{\ttfamily}l p{0.6\textwidth}}
            \toprule
            \rmfamily Code & \rmfamily Example Quotes \\
            \midrule
            INITIAL\_VIS\_COMPREHENSION & ``\emph{...most people when they get home are just relaxing and eating and drinking ...}'' [$P44$] \\
            INFO\_GAIN\_FROM\_TEXT & ``\emph{...a big reason is COVID pandemic...}'' [$P44$] \\
            INITIAL\_TEXT\_COMPREHENSION & ``\emph{...there has been a new all-time (temperature) record from each country each year...}'' [$P20$] \\
            INFO\_GAIN\_FROM\_VIS & ``\emph{...1994 and 2021 are definitely extreme cases...}'' [$P20$] \\
            CONFIRMATION & ``\emph{Yes the chart shows and illustrates the data stated in the text...}'' [$P40$] \\
            CORRECTION & ``\emph{...but they have issued new rules like pitch clock to shorten length of games.}'' [$P12$] \\
            SYNTHESIS & ``\emph{Climate change affects the US differently, including stronger hurricanes in the South and east, drought in the center and wildfires in the West.}'' [$P50$] \\
            \bottomrule
        \end{tabular}
    \end{subtable}
    
    \vspace{1cm}
    
    \begin{subtable}{\textwidth}
        \centering
        \caption{User Experience and Strategy-Related Codes and User Quotes}
        \begin{tabular}{>{\ttfamily}l p{0.6\textwidth}}
            \toprule
            \rmfamily Code & \rmfamily Example Quotes \\
            \midrule
            PREFERENCE\_TEXT\_FIRST & ``\emph{I'd rather read it first. I don't think I like the idea of just seeing charts for this kind of information. It can be confusing for me.}'' [$P24$] \\
            PREFERENCE\_VIS\_FIRST & ``\emph{I prefer seeing a chart first and it conveys information more effectively...}'' [$P38$]. \\
            DIFFERENCE\_VIS\_VS\_TEXT & ``\emph{The differences is different the ways you look at solving a problem or answering a question...}'' [$P23$] \\
            STRATEGY\_VIS\_FIRST & ``\emph{The chart demonstrates the information visually so for me that helps me understand it better...}'' [$P40$] \\
            STRATEGY\_TEXT\_FIRST & ``\emph{I read the text and then imagine what a chart may look like in order to represent that data that is given. Is the data trending up or down?}'' [$P48$] \\
            \bottomrule
        \end{tabular}
    \end{subtable}
\end{table*}

\section{Stimuli}
\label{appd-all-stimuli}

\subsection*{Stimulus 1}

\autoref{fig:ustemp} shows the visual graphic of the US Temperature Record, and below are its corresponding text paragraphs:
\begin{quote}
    \emph{There is temperature data from more than 7,800 weather stations across the United States, most stations have recorded temperature for at least 65 years.
All-time records have been set somewhere in the country every year since at least 1970, but 2021 stands alone when compared with recent years.
Heat waves in 2002 and 2012 brought unprecedented temperatures to hundreds of cities and towns. Like 2021, 2011 broke numerous cold and heat records. But 2021’s extreme temperatures were spread across large areas of the country and surpassed even more records.}
\end{quote}

\subsection*{Stimulus 2}

\autoref{fig:baseball} shows the visual graphic of Average Baseball Game Length, and below are its corresponding text paragraphs:
\begin{quote}
    \emph{Baseball’s future may look a lot like its past.
Nearly two months into the 2023 season, a series of rule changes — including the new pitch clock, enlarged bases and a ban on the infield shift — has translated into a game that evokes the 1980s more than the 2020s.
At 2 hours 39 minutes, the average game is almost half an hour shorter than it was last season, and about the same length as it was in the 1980s.}
\end{quote}

\subsection*{Stimulus 3}

\autoref{fig:cal} shows the visual graphic of the California Electricity Power, and below are its corresponding text paragraphs:
\begin{quote}
    \emph{California draws more electricity from the sun than any other state. It also has a timing problem: Solar power is plentiful during the day but disappears by evening, just as people get home from work and electricity demand spikes. To fill the gap, power companies typically burn more fossil fuels like natural gas.
That’s now changing. Since 2020, California has installed more giant batteries than anywhere in the world apart from China. They can soak up excess solar power during the day and store it for use when it gets dark.
}
\end{quote}

\subsection*{Stimulus 4}

\begin{figure} [htbp]
    \centering
    \includegraphics[width=\linewidth]{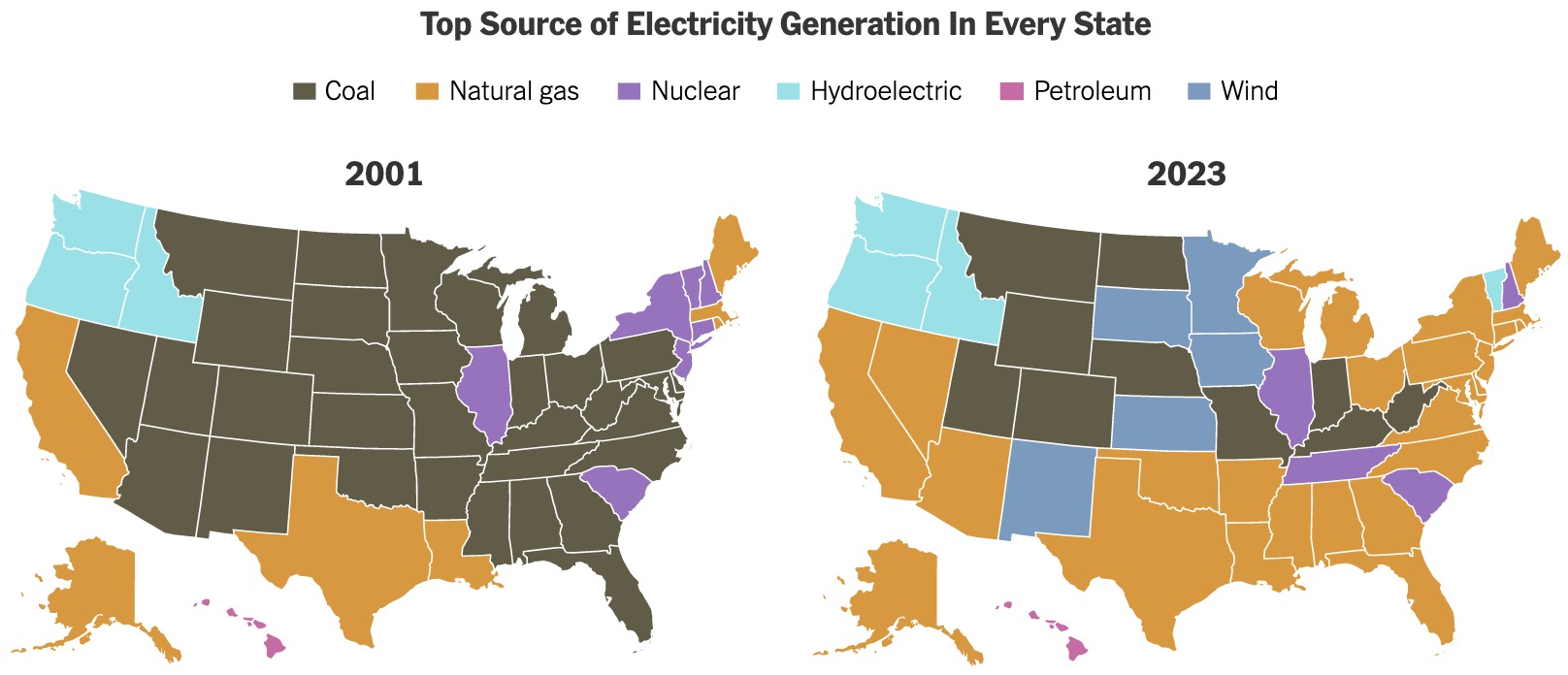}
    \caption{The visual graphic of US Electronic Generation.}
    \label{fig:NYT4}
\end{figure}
\autoref{fig:NYT4} shows the visual graphic of the US Electronic Generation, and below are its corresponding text paragraphs:
\begin{quote}
    \emph{Natural gas surpassed coal as the country’s top source of power in 2016, and renewables like wind and solar have grown quickly to become major players in the U.S. power system.
Fossil fuels still generate the majority of America’s electricity, but the shift from coal to natural gas and renewable power has helped reduce planet-warming carbon dioxide emissions and other harmful pollution.
In 2023, coal was the top electricity fuel in 10 states, down from 32 states in 2001. Natural gas largely took over during that time, but wind also emerged as a leading power source across the Midwest.}
\end{quote}

\subsection*{Stimulus 5}

\autoref{fig:farmwater} shows the visual graphic of the Farming and Underground Water in Kansas, and below are its corresponding text paragraphs:
\begin{quote}
    \emph{There’s little water left to lay down in Wichita County, in Western Kansas. The wells have begun to go dry.
Irrigation can more than double the amount of corn grown per acre. As farms in the area use up the groundwater, corn yields have declined, erasing decades of gains.
As recently as the late 1990s, Wichita County farmers produced 165 to 175 bushels of corn per acre, well above the national average. But it came at a cost, requiring farmers to drain the aquifer in order to irrigate their crops. The area gets less than 20 inches of rain a year, on average, about one-third less than the continental United States as a whole — not nearly enough to replace the water being pumped from the ground.
As farmers ran out of water, they increasingly switched to what’s called dryland farming, relying on rain alone.}
\end{quote}

\subsection*{Stimulus 6}

\begin{figure} [htbp]
    \centering
    \includegraphics[width=\linewidth]{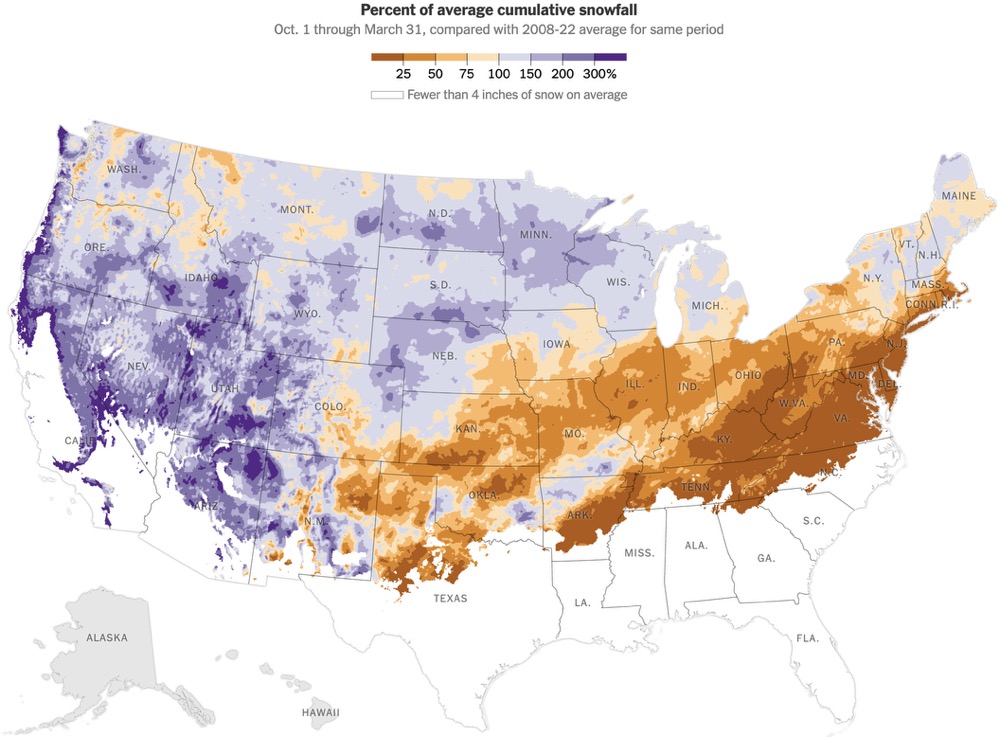}
    \caption{The visual graphic of US Snowfall.}
    \label{fig:NYT6}
\end{figure}
\autoref{fig:NYT6} shows the visual graphic of the US Snowfall, and below are its corresponding text paragraphs:
\begin{quote}
    \emph{2023’s snow season is a tale of two starkly different winters: A cold and snowy one in the West, and a warm and relatively snowless one in the East.
The Western United States received a lot more snow than usual this season, much of it unleashed by punishing storms that battered California especially hard throughout the winter. Parts of the Eastern half of the country, however, saw much less snow than normal amid unusually warm winter temperatures.
It’s not uncommon for the two coasts to experience opposing weather conditions. This can happen when the jet stream, a band of winds that blow from west to east around the planet, starts to meander into a wavelike pattern. This wind pattern results in cooler conditions where it dips southward and warmer conditions where it arcs northward.
}
\end{quote}

\subsection*{Stimulus 7}

\begin{figure} [htbp]
    \centering
    \includegraphics[width=\linewidth]{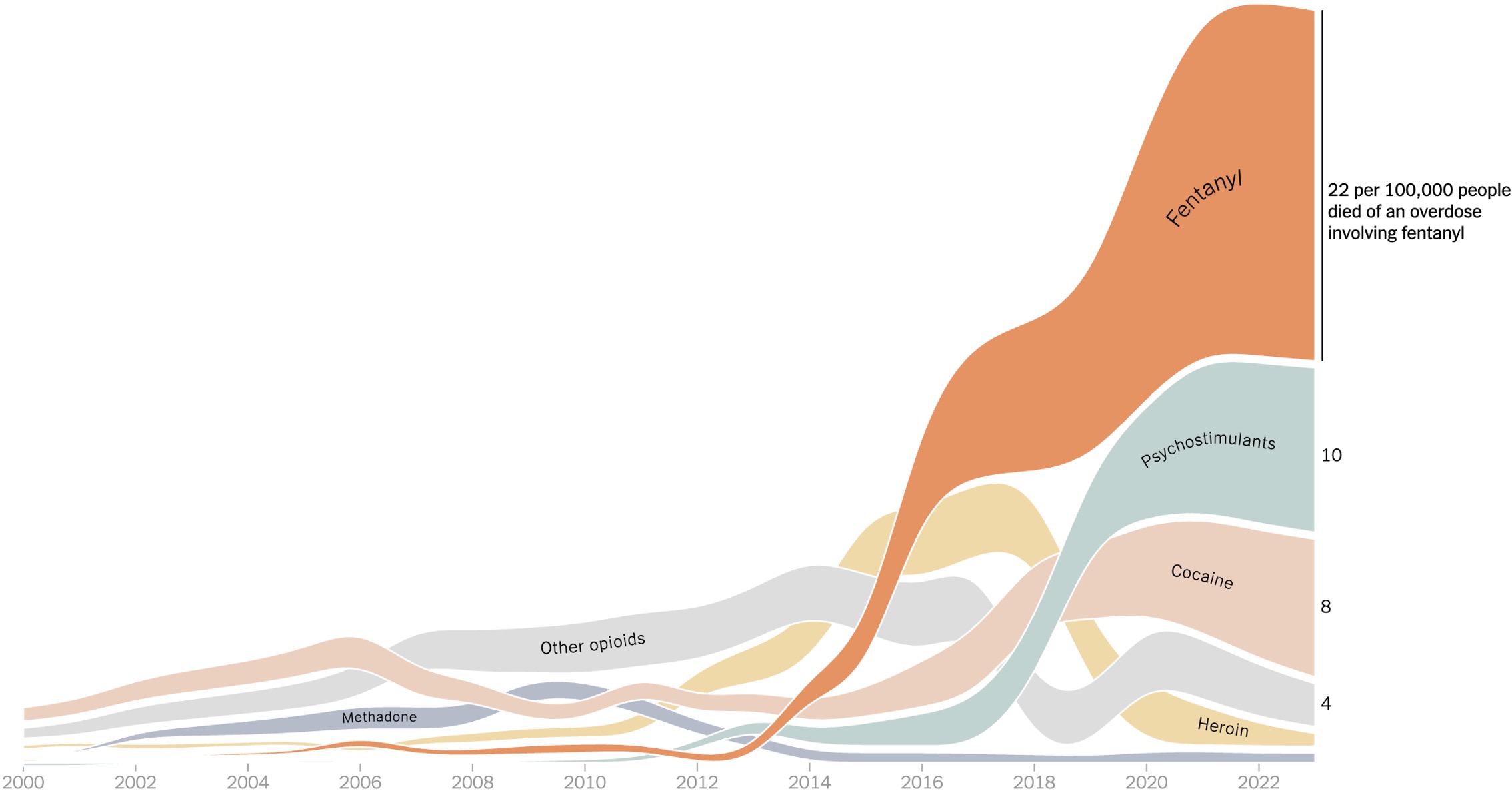}
    \caption{The visual graphic of Deaths from Drug Mixtures.}
    \label{fig:NYT7}
\end{figure}
\autoref{fig:NYT7} shows the visual graphic of the Deaths from Drug Mixtures, and below are its corresponding text paragraphs:
\begin{quote}
    \emph{Last year over 70,000 Americans died from taking drug mixtures that contained fentanyl or other synthetic opioids. The good news is that recent data suggests a decline in overdose deaths, the first significant drop in decades. But this is not a uniform trend across the nation. To understand this disparity, it’s important to examine how we got here.
Today’s crisis is often described as a series of waves. But if you look at the data, it was more like a couple of breakers followed by a tsunami. First, prescription opioid fatalities rose. Then heroin deaths surged. And finally, illicitly manufactured fentanyl overtook all that preceded it.
}
\end{quote}

\subsection*{Stimulus 8}

\begin{figure} [htbp]
    \centering
    \includegraphics[width=\linewidth]{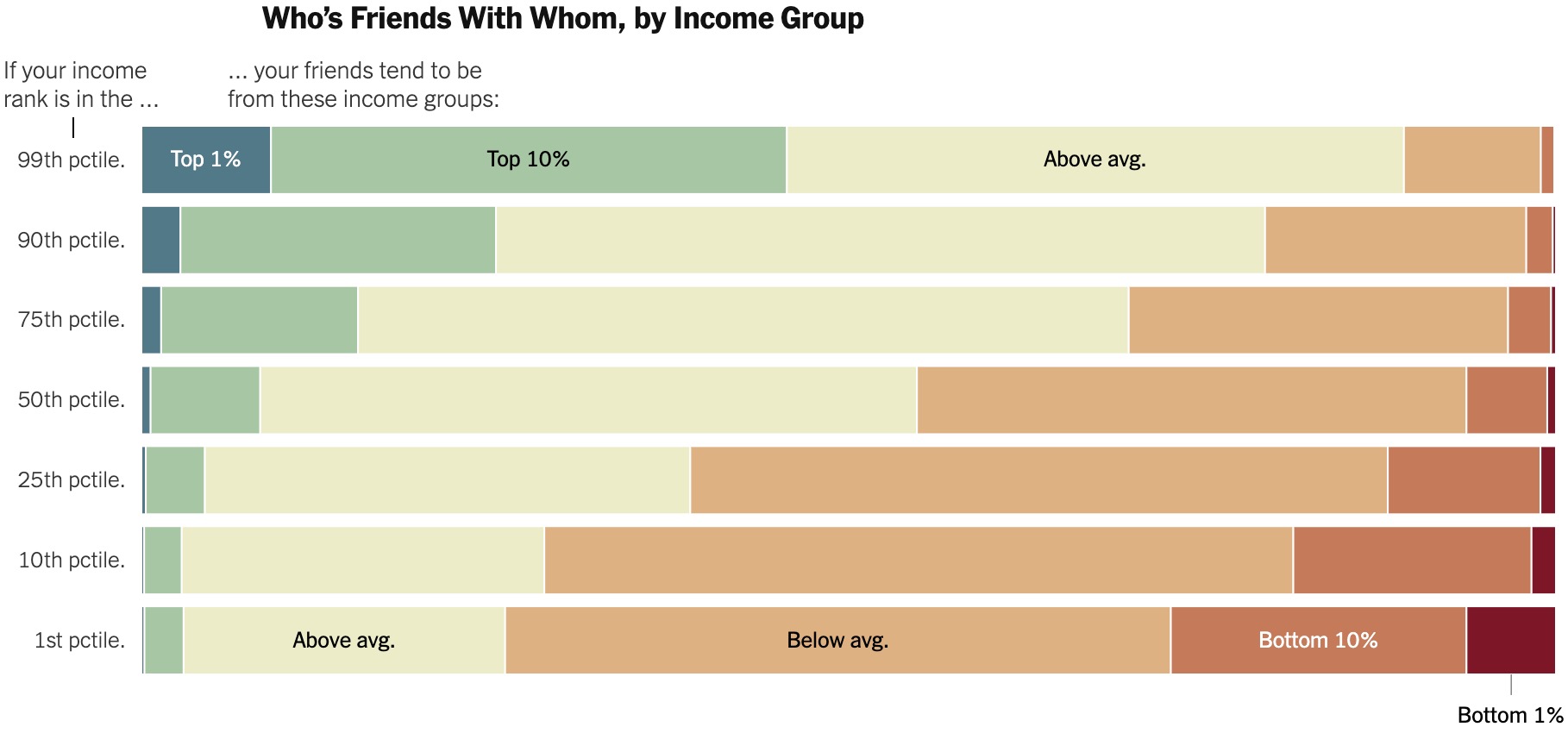}
    \caption{The visual graphic of Friends and Income.}
    \label{fig:NYT8}
\end{figure}
\autoref{fig:NYT8} shows the visual graphic of the Friends and Income, and below are its corresponding text paragraphs:
\begin{quote}
    \emph{The pressure that parents feel to try to give their kids a competitive advantage is amplified when society is unequal and there’s more to be lost. Our society is structured in ways that discourage these cross-class friendships from happening, and many parents are making choices about where to live and what extracurriculars to put their kids into that make those friendships less likely to happen.
As a result, rich people have mostly rich friends, and poor people have mostly poor friends.
Low-income people are far more likely than high-income people to make friends in their neighborhoods. But in poorer areas, there are fewer rich people nearby to befriend.
}
\end{quote}

\subsection*{Stimulus 9}

\begin{figure} [h]
    \centering
    \includegraphics[width=\linewidth]{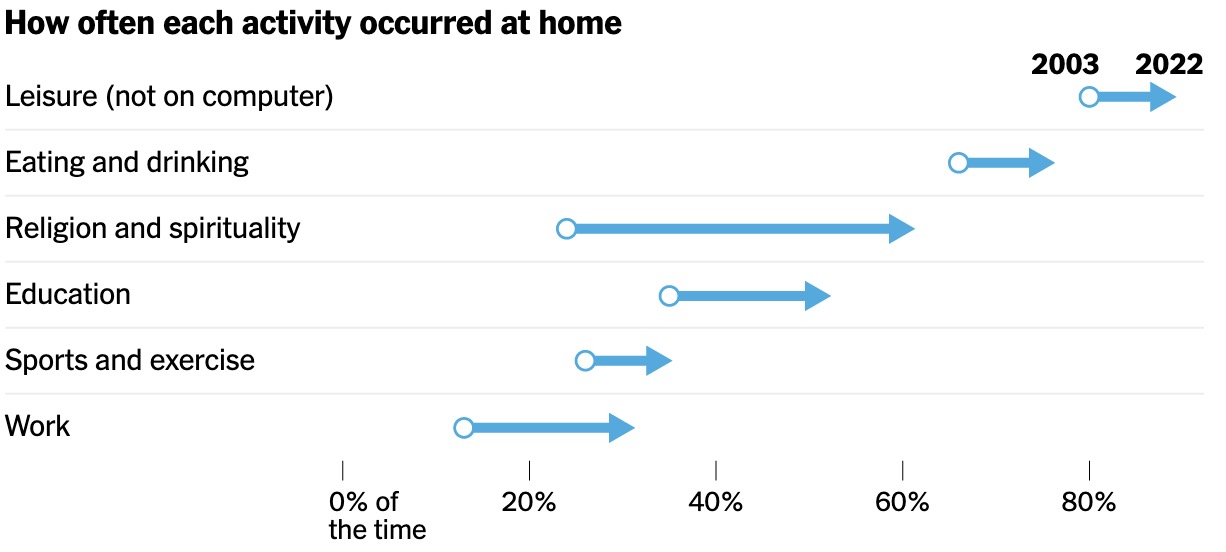}
    \caption{The visual graphic of Activities At Home.}
    \label{fig:NYT9}
\end{figure}
\autoref{fig:NYT9} shows the visual graphic of the Activities At Home, and below are its corresponding text paragraphs:
\begin{quote}
    \emph{Americans’ time spent at home increased by 1 hour 39 minutes a day, or 10 percent, from 2003 through 2022.
The rise in working at home during the pandemic has been a big chunk of that, taking up 29 percent of all work activity in 2022. Other activities followed a similar track — a gradual rise over years followed by a spike during the pandemic that was still felt sharply in the first two months of 2022.
As of 2022, time that Americans had once spent outside the home participating in activities like education, eating and drinking, had, to some extent, moved into the home. The largest shift occurred with religious activities: 59 percent occurred at home in 2022, up from 24 percent in 2003.}
\end{quote}

\subsection*{Stimulus 10}

\begin{figure} [htbp]
    \centering
    \includegraphics[width=\linewidth]{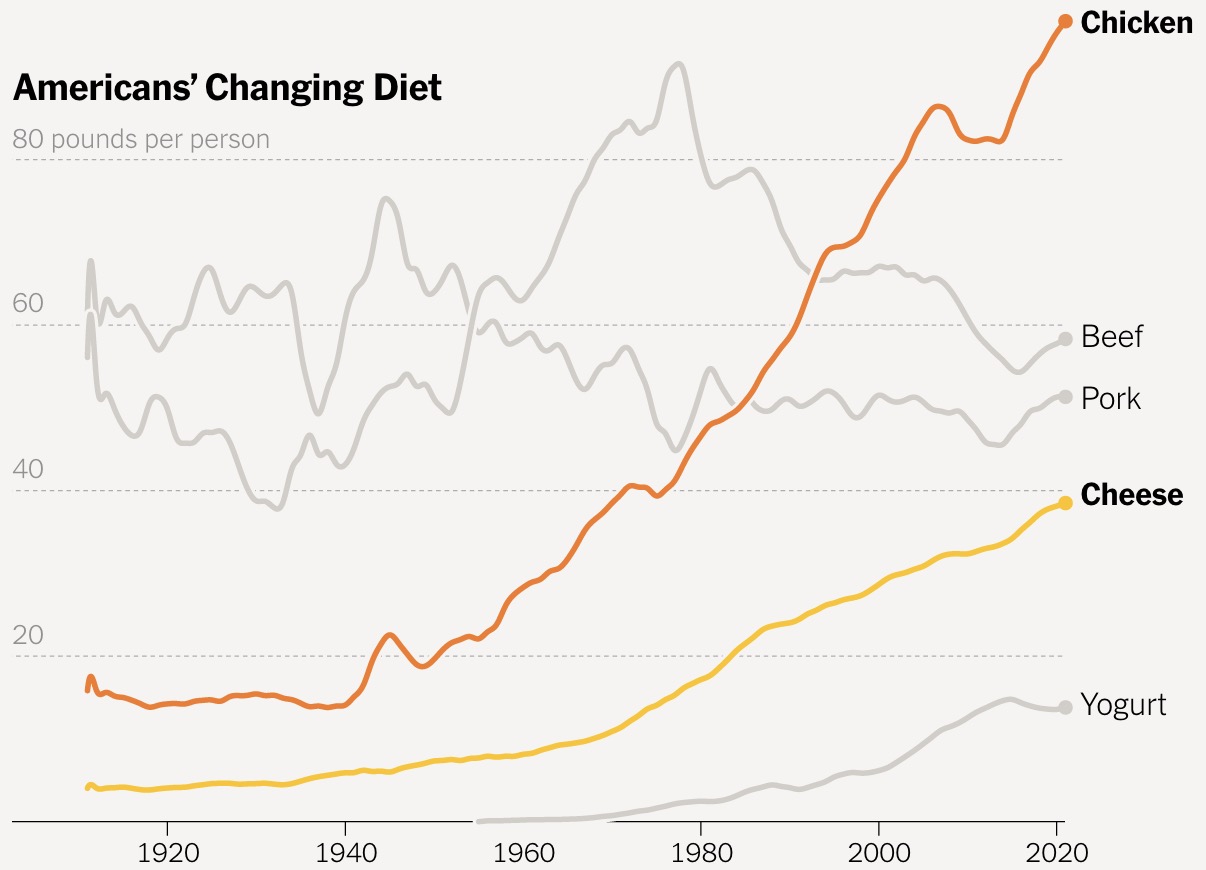}
    \caption{The visual graphic of Americans' Changing Diet.}
    \label{fig:NYT10}
\end{figure}
\autoref{fig:NYT10} shows the visual graphic of the Americans' Changing Diet, and below are its corresponding text paragraphs:
\begin{quote}
    \emph{America's striking dietary shift in recent decades, toward far more chicken and cheese, has not only contributed to concerns about American health but has taken a major, undocumented toll on underground water supplies.
These transformations are tied to the changing American diet. Since the early 1980s, America's per-person cheese consumption has doubled, largely in the form of mozzarella-covered pizza pies. And last year, for the first time, the average American ate 100 pounds of chicken, twice the amount 40 years ago.}
\end{quote}

\subsection*{Stimulus 11}

\begin{figure} [htbp]
    \centering
    \includegraphics[width=\linewidth]{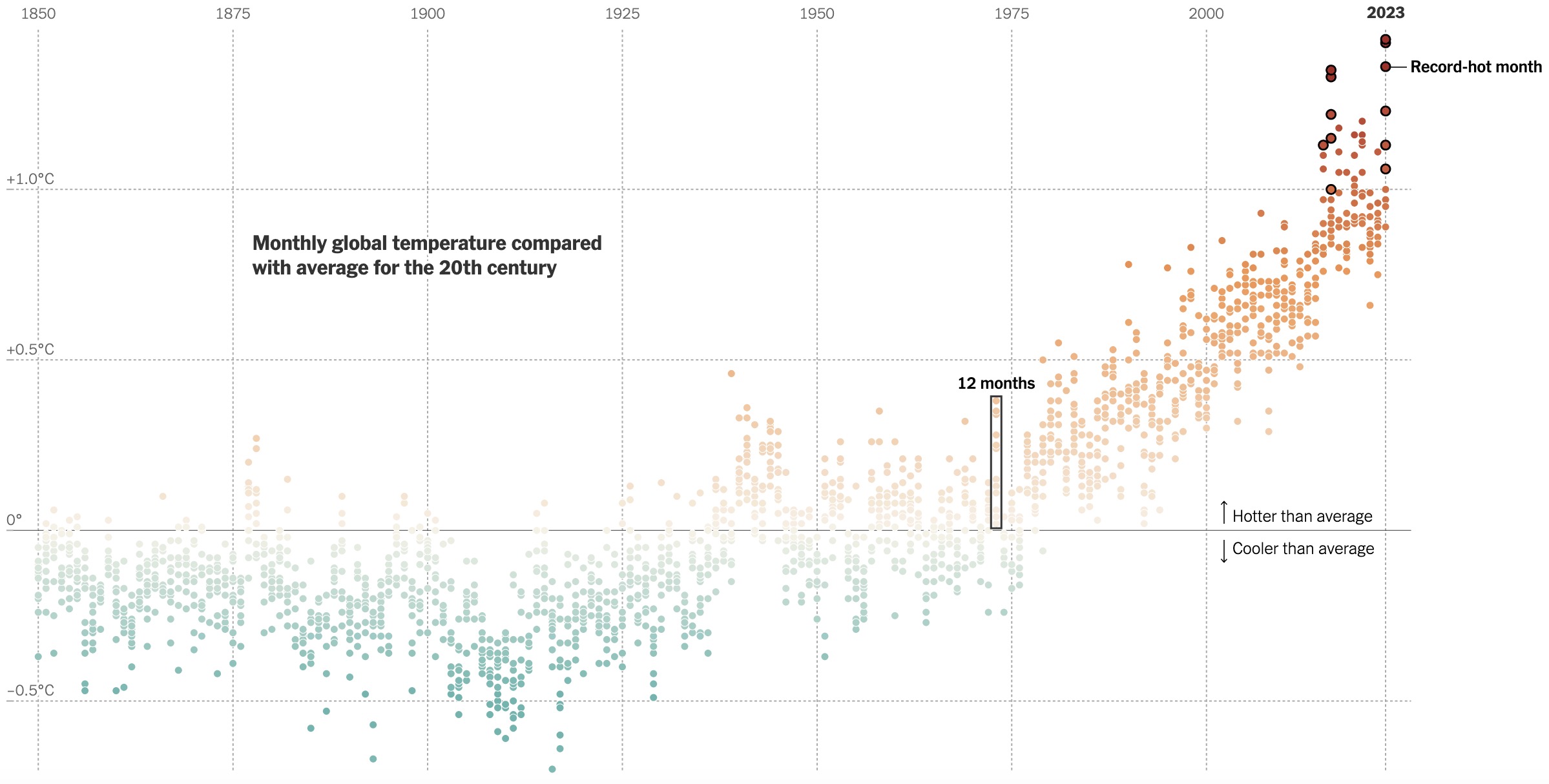}
    \caption{The visual graphic of Global Temperature.}
    \label{fig:NYT11}
\end{figure}
\autoref{fig:NYT11} shows the visual graphic of the Global Temperature, and below are its corresponding text paragraphs:
\begin{quote}
    \emph{2023 is the Earth’s warmest year in the past 174 years, and very likely the past 125,000. Global temperatures started blowing past records midyear and didn’t stop. First, June was the planet’s warmest June on record. Then, July was the warmest July. And so on, all the way through December.
Averaged across last year, temperatures worldwide were 1.48 degrees Celsius, or 2.66 Fahrenheit, higher than they were in the second half of the 19th century. That is warmer by a sizable margin than 2016, the previous hottest year.
One hypothesis, perhaps the most troubling, is that the planet’s warming is accelerating, that the effects of climate change are barreling our way more quickly than before. 
}
\end{quote}

\subsection*{Stimulus 12}

\begin{figure} [htbp]
    \centering
    \includegraphics[width=\linewidth]{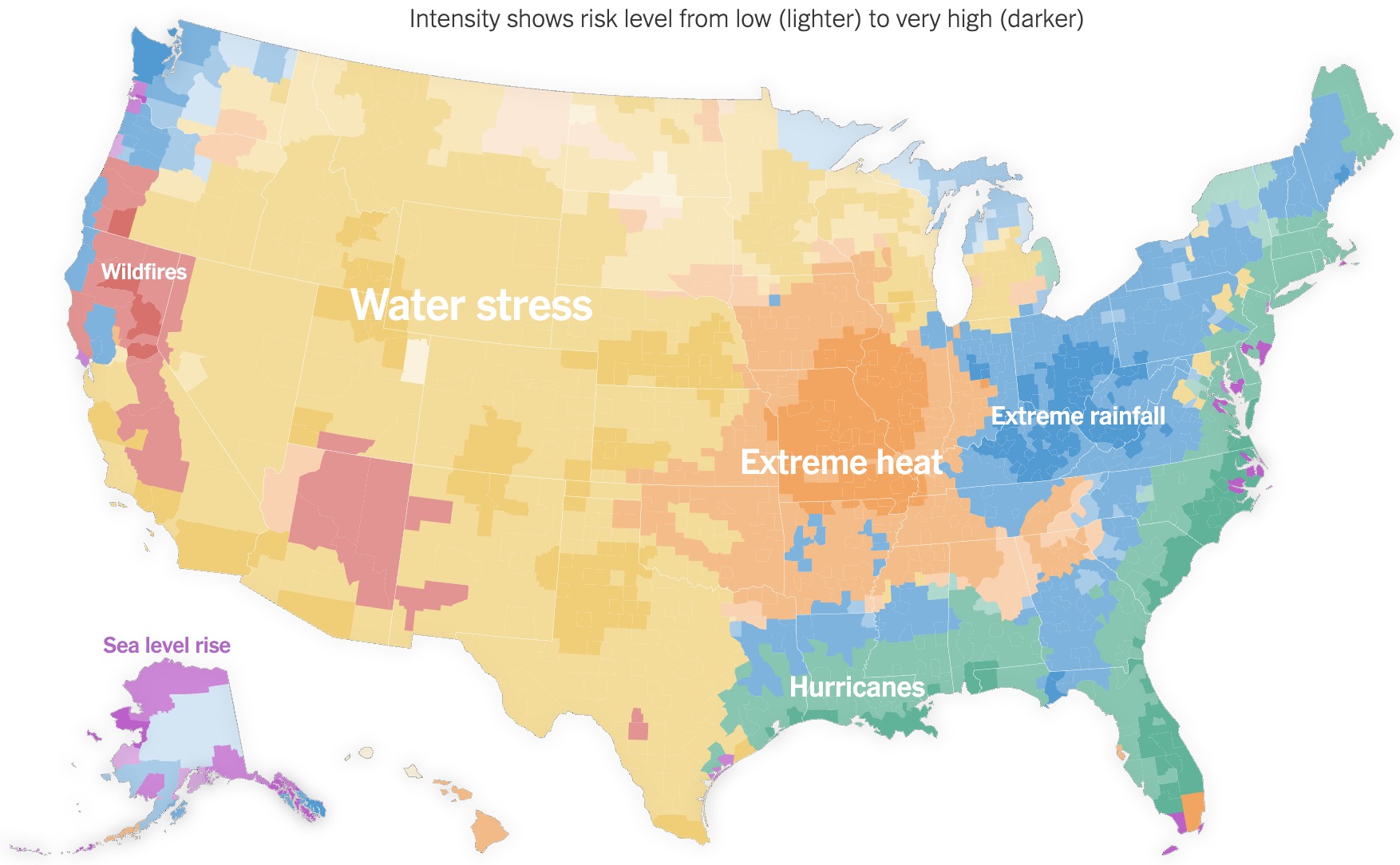}
    \caption{The visual graphic of Disaster and Climate Risks.}
    \label{fig:NYT12}
\end{figure}
\autoref{fig:NYT12} shows the visual graphic of the Disaster and Climate Risks, and below are its corresponding text paragraphs:
\begin{quote}
    \emph{For most of us, climate change can feel like an amorphous threat — with the greatest dangers lingering ominously in the future and the solutions frustratingly out of reach.
Thinking this way: looking at the most significant climate threat unfolding in your own backyard. Transforms the West Coast’s raging wildfires into “climate fires.” The Gulf Coast wouldn’t live under the annual threat of floods but of “climate floods.” Those are caused by ever more severe “climate hurricanes.” The Midwest suffers its own “climate droughts,” which threaten water supplies and endanger crops.
Every single county has some sort of climate threat that’s either emerged and is doing some damage right now or is going to emerge.}
\end{quote}

\end{document}